\documentclass[useAMS,usenatbib]{mn2e}
\pdfoutput=1
\usepackage{aas_macros}
\usepackage{amsmath}
\usepackage{color}
\usepackage{epsfig}

\newcommand{\Msol}{\rm\,M_{\odot}}
\newcommand{\Mstellar}{\rm\,M_{*}} 
\newcommand{\Mhalo}{\rm\,M_{halo}}
\newcommand{\Gyr}{\rm\,Gyr} 
\newcommand{\Myr}{\rm\,Myr}
\newcommand{\yr}{\rm\,yr} 
\newcommand{\Msolyr}{\rm\,M_{\odot}\,yr^{-1}}
\newcommand{\Mpc}{\rm\,Mpc} 

\newcommand{\sfr}{\rm\,SFR} 
\newcommand{\ssfr}{\rm\,SSFR}

\newcommand{\kmsMpc}{\rm\,km\,s^{-1}\,Mpc^{-1}}

\newcommand{\MB}{\ensuremath{M_{B}}}

\newcommand{\zsat}{\rm\,z_{sat}}
\newcommand{\tsat}{\rm\,t_{sat}}
\newcommand{\bt}{\rm\,B/T}

\newcommand{\tausat}{\rm\,\tau_{sat}}
\newcommand{\lesssim}{\rm\,\la}
\newcommand{\gtrsim}{\rm\,\ga}

\newcommand{\subfind}{\textsc{subfind}}
\newcommand{\morgana}{\textsc{morgana}}
\newcommand{\morganabf}{\bf \textsc{MORGANA}}
\newcommand{\pinocchio}{\textsc{pinocchio}}
\newcommand{\hop}{\textsc{HOP09}}

\defcitealias{Wilman12}{WE12}
\defcitealias{deLucia11}{DL11}
\defcitealias{WDL08}{WDL08}

\title[Hierarchical Origins of Morphology]
{The Hierarchical Origins of Observed Galaxy Morphology}

\author[D.~J.~Wilman, et al.]
{David~J.~Wilman$^{1,2,*}$,~Fabio Fontanot$^{3,4,5}$,~Gabriella De Lucia$^{5}$,~Peter Erwin$^{1,2}$,
\newauthor
Pierluigi Monaco$^{5,6}$\\
       $^1$Max-Planck-Insitut f\"{u}r Extraterrestrische Physik, Giessenbachstrasse, 85748 Garching, Germany.\\
       $^2$Universit\"{a}ts-Sternwarte M\"{u}nchen, Scheinerstrasse 1, 81679 M\"{u}nchen, Germany.\\
       $^3$HITS-Heidelberger Institut f\"ur Theoretische Studien, Schloss-Wolfsbrunnenweg 35, 69118 Heidelberg, Germany.\\
       $^4$Institut f\"ur Theoretische Physik, Philosophenweg, 16, 69120, Heidelberg, Germany.\\
       $^5$INAF - Astronomical Observatory of Trieste, via G.B. Tiepolo 11, I-34143 Trieste, Italy.\\
       $^6$Dipartimento di Astronomia, Universit\`a di Trieste, via G.B. Tiepolo 11, I-34131 Trieste,Italy.\\
       $^*$email: dwilman@mpe.mpg.de}

\begin{document}

\maketitle

\begin{abstract}

  Galaxies grow primarily via accretion-driven star formation in discs
  and merger-driven growth of bulges. These processes are implicit in
  semi-analytical models of galaxy formation, with bulge growth in
  particular relating directly to the hierarchical build-up of halos
  and their galaxies.  In this paper, we consider several
  implementations of two semi-analytical models.  Focusing on
  implementations in which bulges are formed during mergers only, we
  examine the fractions of elliptical galaxies and both passive and
  star-forming disk galaxies as functions of stellar and halo mass,
  for central and satellite systems. This is compared to an
  observational cross-matched SDSS $+$ RC3 (SDSSRC3) $z \sim 0$ sample
  of galaxies with accurate visual morphological classifications and
  $\Mstellar>10^{10.5}\Msol$. The models qualitatively reproduce the
  observed increase of elliptical fraction with stellar mass, and with
  halo mass for central galaxies, supporting the idea that observed
  ellipticals form during major mergers. However, the overall
  elliptical fraction produced by the models is much too high compared
  with the $z \sim 0$ data. Since the ``passive'' -- i.e.
  non-star-forming -- fractions are approximately reproduced, and
  since the fraction which are star-forming \textit{disc} galaxies is
  also reproduced, the problem is that the models overproduce
  ellipticals at the expense of \textit{passive} S0 and spiral
  galaxies. Bulge-growth implementations (tuned to reproduce
  simulations) which allow the survival of residual discs in major
  mergers still destroy too much of the disc. Increasing the lifetime
  of satellites, or allowing significant disc regrowth around merger
  remnants, merely increases the fraction of star-forming disc
  galaxies. Instead, it seems necessary to reduce the mass ratios of
  merging galaxies, so that most mergers produce modest bulge growth
  in disc-galaxy remnants instead of ellipticals. This could be a
  natural consequence of tidal stripping of stars from infalling
  satellite galaxies, a process not considered in our models. However,
  a high efficiency of quenching during and/or subsequent to minor
  mergers is still required to keep the passive fraction high.

\end{abstract}


\section{Introduction}\label{sec:intro}


Galaxy morphology has long been known to correlate strongly with local
environment
\citep{Melnick77,Dressler80,Postman84,Wilman09,Bamford09,Wilman12} and
stellar mass \citep{Bamford09,Vulcani11b,Wilman12}. This relates both
to the structural components of galaxies \citep[bulge, disc, spiral
arms, bar,][]{Hubble26} and to star formation \citep[correlated with
spiral structure, with low or zero star formation rates in early type
galaxies,][]{Sandage78,Sellwood11}.


Simulations show that elliptical galaxies and classical
(pressure-supported) bulges in galaxies can be formed when galaxies
merge \citep[e.g.][]{Barnes88,Springel05}.  In a hierarchical, cold
dark matter dominated universe, dark matter halos build up through
regular mergers and smooth accretion \citep{Lacey93,Genel10,Wang11}.
Galaxies which have grown at the centre of merging halos will then
orbit within the conglomerate global potential, experience tidal
stripping, and eventually, through dynamical friction, sink towards the
bottom of the remnant halo where they will merge. At this point a
classical bulge or elliptical galaxy may form through violent
relaxation \citep{LyndenBell67}. The implication is that elliptical and
classical bulge formation relates directly to the hierarchical growth
of structure in the universe. As such, the abundance, mass and
environmental dependence of ellipticals and classical bulges can be
quantified in the context of dark matter simulations, provided halos
and their merger trees can be realistically populated with galaxies.


Semi-analytic models of galaxy formation
\citep[e.g.][]{White91,Bower06,Monaco07,deLucia07} provide a
self-consistent, physically motivated framework to understand the
hierarchical evolution of galaxies, and connect this to their
morphological evolution.  Physically motivated prescriptions for
cooling and heating of gas and its conversion to stars are applied to
halo merger trees, tracking the star formation and merger-related
growth of galaxies \citep[See][for a review of these
techniques]{Baugh06}. This framework provides an ideal way to examine
the resultant distribution of galaxy properties such as stellar mass
($\Mstellar$), bulge to total ratio ($\bt$) and star formation rates
(SFR), as well as their dependance on each other, and on environment at
a given redshift slice. Comparison with observations then allows one to
place constraints on the physical processes governing the baryonic
assembly and merger-induced (classical) bulge growth in galaxies.

This paper presents such a comparison.  We have considered several
bulge formation implementations which have been applied to two models
of galaxy formation \citep[][hereafter
\citetalias{deLucia11}]{deLucia11}. As described by
\citetalias{deLucia11}, these models provide us with an ideal tool to
examine how dynamical friction, bulge growth and star formation
histories bring about a model galaxy's $\bt$ and SFR at $z=0$, and
their dependence on stellar and halo mass. Our models have also been
applied to compare the predicted abundance of bulgeless galaxies to
their observed abundance \citep{Fontanot11} and the differences in
merger history of elliptical-rich clusters to that in elliptical-poor
clusters \citep{deLucia12}. In \citetalias{deLucia11} we find that the
alternate method of forming bulges via the instability of discs is
currently ill-constrained in models. This motivates our choice in this
paper to emphasize model implementations in which the disc instability
mode is switched off, and bulges only grow during mergers.

In Section~\ref{sec:models} we present our models, outlining the key
physical processes important for this analysis. We have also
constructed a local sample of galaxies with visual morphological
classifications \citep[][hereon \citetalias{Wilman12}]{Wilman12}, and
with well defined stellar masses and environments.  This sample is
presented in Section~\ref{sec:sdssrc3}.  In Section~\ref{sec:btmass} we
examine the distribution of model galaxies in $\bt$ versus $\Mstellar$,
compared with recent observations.  Then in
Section~\ref{sec:ellipticals} we examine the dependence of elliptical
fraction, separately for central and satellite galaxies, on stellar and
halo mass. In Section~\ref{sec:pdisc} we examine the fraction of
star-forming disc, and passive disc galaxies in the same way, and in
Section~\ref{sec:passive} we examine the total passive fraction. The
results presented in Section~\ref{sec:results} motivate our
interpretation of how observed galaxy morphology and star formation are
imprinted by hierarchical growth, and of the remaining deficiencies in
the physical prescriptions of our models. This interpretation is
presented in Section~\ref{sec:interpretation}.  We summarize and
discuss the prospects for improving our models in
Section~\ref{sec:conclusions}.

\section{Models}\label{sec:models}

We examine the origin of galaxy morphology in the context of two
independently developed semi-analytic models of galaxy formation. These
are the Munich model, as implemented by \citet{deLucia07} and
generalized to WMAP3 cosmological parameters, as discussed by
\citet{WDL08} (we refer the reader to this model as
\citetalias{WDL08}), and the \morgana\ model \citep{Monaco07} adapted
to a WMAP3 cosmology by \citet{LoFaro09}.  Comparisons between these
models have been presented in \citet{Fontanot09}, \citet{Fontanot11b},
\citet{deLucia10} and \citetalias{deLucia11}, and we refer the reader
to these papers for more details.

\citetalias{WDL08} assumes a cosmology with $H_0 = 74.3 \kmsMpc$,
$\Omega_m = 0.226$, $\sigma_8 = 0.722$, $n = 0.947$ and $\Lambda_0 = 0.774$
while \morgana\ assumes $H_0 = 72 \kmsMpc$, $\Omega_m = 0.24$,
$\sigma_8 = 0.8$, $n = 0.96$ and $\Lambda_0 = 0.77$.  For the purposes of
this paper, these small differences in cosmology have a negligible
effect on results.

In this Section we consider the aspects most pertinent to galaxy
morphology and therefore the results presented in this paper.  These
include the bulge formation implementations
(Section~\ref{sec:bulgeformation}) and survival time of satellites
(Section~\ref{sec:satsurvivaltime}). We also briefly overview key
results from \citetalias{deLucia11}, relevant to our analysis
(Section~\ref{sec:dl11}).

\subsection{Bulge Formation Implementations}\label{sec:bulgeformation}

Models assume gas to cool onto a disc, which then forms a stellar disc
via star formation. The formation of bulges requires the loss of
angular momentum, which happens either when galaxies merge or when
discs become unstable.

\citetalias{deLucia11} (to which we refer for further details)
presented three implementations for bulge formation, each applied to
both \citetalias{WDL08} and \morgana\ models. The {\it standard}
implementations are those applied by default to both models, in which
bulges form via both disc instabilities and galaxy mergers. The {\it
  pure mergers} implementations exclude disc instabilities so that we
can isolate the effects of galaxy mergers.

In a major merger (baryonic mass ratio $\mu>0.3$) the standard and
pure-merger implementations put all pre-existing stars into the bulge of
the remnant galaxy, thus forming an elliptical with $\bt=1$ {\it by
  definition}. The cold gas from both progenitors fuels a starburst
which adds to this bulge mass. The galaxy can then grow a new disc at
later times from gas accreted onto the remnant.

Disc instabilities are treated differently in the two models.  When the
instability criterium is met (see \citetalias{deLucia11}),
\citetalias{WDL08} transfers just enough stellar mass to the bulge to
restore stability, while \morgana\ transfers half of the baryonic mass
to the bulge. This leads to a much stronger role for disc instabilities
in the formation of massive bulges in \morgana\ than in
\citetalias{WDL08}, with the standard bulge growth implementations.

Both models include the option of using the {\it \hop} implementation,
which are a modification of the pure-mergers case, tuned to the results
of idealized merger simulations from \citet{Hopkins09}.  The two major
differences between this and the standard pure-merger approach relate
to the treatment of stars and gas in a major merger.  A fraction
$1-\mu$ of the stellar disc from the more massive (primary) progenitor
survives the merger. Also, only a fraction of the cold gas from the
primary progenitor goes into the starburst; the remaining gas is
retained by the disc of the remnant where normal star formation
subsequently takes place. The starbursting gas fraction decreases with
increasing total gas fraction and increases with increasing mass ratio.
These implementations tend to leave residual stellar and gas discs
which would be completely destroyed during a major merger in the
standard or pure-merger prescriptions.

\subsection{Satellite survival time}\label{sec:satsurvivaltime}

Once a smaller halo is accreted onto a larger one it becomes a subhalo,
and the galaxy at its centre is now considered a satellite of the
parent halo. Dynamical friction draws the subhalo towards the parent
halo's core. As the subhalo moves into regions of higher surrounding
density, tidal stripping becomes more effective. This reduces the
subhalo mass, and increases the efficiency of dynamical friction: these
two effects are intertwined. Eventually the satellite galaxy reaches
the centre of the parent halo, where it merges with the central galaxy.
We denote the time from accretion of the subhalo (when the galaxy
becomes a satellite) until the galaxies merge the {\it satellite
  survival time} or $\tausat$.

This timescale is clearly important: shorter timescales will lead to
more mergers and fewer satellites, and less time for satellite-specific
processes to act. However, it varies significantly between different
numerical determinations.

\citetalias{WDL08} defines merger trees for subhaloes, that are
identified in the N-body simulation using the algorithm \subfind\
\citep{Springel01}. The merger trees are then constructed using a
dedicated software that is the same developed to analyse the Millennium
Simulation.  Subhaloes are tracked until they are tidally stripped to a
point at which they can no longer be resolved. The semi-analytic model
then assigns a residual survival time to the ``orphaned'' satellite
galaxy according to a dynamical friction formula.

In \morgana, the code \pinocchio\ \citep{Monaco02,Taffoni02} which is
based on Lagrangian perturbation theory, is used to construct mass
assemly histores of dark matter halos which are then populated using
the \morgana\ semi-analytic model. As \pinocchio\ does not follow the
evolution of substructures, a (slightly updated) version of the fitting
formulae provided by \citet{Taffoni03} are applied to compute
$\tausat$. This interpolates between the cases of a 'live satellite'
(in which the subhalo experiences significant mass loss) and a 'rigid
satellite' (with no mass loss). In the version of \morgana\ used in
this paper, stellar stripping is {\it not} included.  Initial orbital
parameters for satellite galaxies are randomly extracted from suitable
distributions. In contrast to \citetalias{WDL08}, the clock for
satellite survival (and its orbit) is reset after every DM halo major
merger.

\citet{deLucia10} examined the implied dynamical friction timescales of
these and other models as a function of the progenitor mass ratio
$\mu$, and found them to be widely variant. In particular, massive
satellites with $\mu>0.1$ survive for an order of magnitude longer by
\citetalias{WDL08} than for \morgana.

In \citetalias{deLucia11}, we examined the effects of a longer
dynamical friction time for \morgana\ by using the {\it longer
  $\tausat$ } dynamical friction timescale from \citetalias{WDL08}. We
shall also consider this adaptation of the \morgana\ model in this
paper.  However, we emphasize that this version of the \morgana\ model
has not been recalibrated to fit other observables, and so these
results should be interpreted with caution.  We also stress that even
when adopting the same formula used in \citetalias{WDL08} in \morgana,
satellite survival times will not be identical because of different
assumptions adopted in the two models.  For details, we refer the
reader to \citet{deLucia10}.

\subsection{\citetalias{deLucia11}: A Summary of Results}\label{sec:dl11}

In \citetalias{deLucia11} we found a strong correlation between galaxy
bulge fraction ($\bt$) and stellar mass, and between bulge fraction and
halo mass for central galaxies, such that central galaxies of
$\Mhalo\gtrsim10^{13}\Msol$ halos are bulge-dominated. This is a direct
consequence of the richer merger history for more massive galaxies
which live at the centre of a more massive halo.

We examined the different channels for bulge growth, and found that
major mergers dominate bulge growth of $\Mstellar\lesssim10^{10}\Msol$
galaxies, while minor mergers produce comparible bulge mass in more
massive galaxies. However, the vast majority of bulge-dominated
($\bt>0.9$) galaxies acquired their high bulge fractions through major
mergers.  In the standard implementations for bulge growth, disc
instabilities dominate the formation of bulges in intermediate mass
galaxies ($\sim 10^{10}-10^{11}\Msol$) and can also lead the the
formation of bulge-dominated galaxies at high redshift in the \morgana\
model.

In our models, bulge-dominated galaxies can grow a new stellar disc:
hot gas cools to form a new cold gas disc which then forms new stars.
\citetalias{deLucia11} showed that this disc regrowth rate is highest
in intermediate mass galaxies ($\sim 10^{10}-10^{11}\Msol$) and
increases with redshift. The fraction of bulge-dominated central
galaxies regrowing a disc depends on the model. The fraction of
\morgana\ central galaxies experiencing regrowth increases with time to
almost $100\%$ at $z=0$, while the corresponding fraction for
\citetalias{WDL08} decreases to $\lesssim 50\%$ at $z=0$ for $\Mstellar
\geq 10^{10}\Msol$, with lower fractions at high mass.  Although large
numbers of galaxies experience disc regrowth, the rate of regrowth is
modest, particularly at low redshift.  Both models implement a
radio-mode AGN feedback, but this is particularly efficient in the
\citetalias{WDL08} model. This suppresses the cooling of the hot gas
with an efficiency which is a strong function of halo mass. Thus the
most massive galaxies which live at the centre of the most massive
halos, especially at low redshift, experience a stronger feedback and
less regrowth of their discs.

\section{The SDSSRC3 sample}\label{sec:sdssrc3}

While there now exist large samples of classified galaxies in the local
Universe, our goals require the identification of galaxies with
significant discs: this means it is essential to separate elliptical
from S0 morphological types. This division was not considered, for
example, in the initial Galaxy Zoo classification scheme
\citep{Lintott08,Bamford09}.

In \citetalias{Wilman12} \citep{Wilman12} we constructed a sample of
$z\sim0$ galaxies with robust morphological classification based upon
the New York University Value Added Galaxy Catalog
\citep[NYU-VAGC,][]{Blanton05} who matched the SDSS DR4 \citep[Sloan
Digital Sky Survey Data Release 4, ][]{SDSSDR4} to the Third Reference
Catalog of Bright Galaxies \citep[RC3,][]{RC3}.  This provides
Hubble-type morphological classifications for 1194 galaxies with
B-magnitudes $B\leq16$, including 165 galaxies which we have
re-classified and 55 which are classified for the first time (based on
SDSS imaging).  As described by \citetalias{Wilman12}, we weight
galaxies to correct for the RC3 selection bias as a function of
$B$-band magnitude and the radius containing 50\% of the Petrosian flux
in r-band, and to correct for Malmquist bias ($V/V_{\rm max}$).

We calculated stellar masses for each galaxy using the color-based
mass-to-light ($M/L$) ratios of \citet{Zibetti09}, using SDSS $g - i$
colors and $i$-band absolute magnitudes (including the necessary
k-corrections).  Stellar masses have been corrected for
over-subtraction of the SDSS sky background, which is significant for
galaxies larger than $r_{50}\sim 10\arcsec$, where $r_{50}$ is the
radius containing half the $r$-band Petrosian flux \citep{Blanton11}.

In \citetalias{Wilman12} we examine the galaxy population limited to
$\MB\leq-19$, corresponding to a red galaxy with a stellar mass
$\Mstellar \gtrsim 10^{10.5}\Msol$. For this paper we have checked that
for galaxies $\Mstellar \geq 10^{10.5}\Msol$, the morphological
composition is almost identical with or without an additional cut in
luminosity at $\MB\leq-19$. To keep interpretation simple, we have
chosen to apply just the cut in stellar mass. This leaves us with 854
galaxies in the sample.

We take halo masses from the {\it Sample II} group catalog of
\citet{Yang07} constructed from the SDSS-DR4.  We refer the reader to
\citet{Yang07} and \citetalias{Wilman12} for a full description of this
catalogue and its application to our sample.  In brief: a
friends-of-friends linking algorithm is used to assign galaxies to
groups, which are then assigned halo masses based upon the rank order
in terms of the group total stellar mass of all galaxies brighter than
an evolution and k-corrected r-band absolute magnitude of $-19.5$. An
isolated galaxy with this limiting luminosity is assigned a halo mass
of $\Mhalo=10^{11.635}\Msol$, which is therefore the \citet{Yang07}
halo mass limit, while an isolated early-type galaxy with a stellar
mass $\Mstellar = 10^{10.5}\Msol$ has a halo mass of $\Mhalo \sim
10^{11.75}\Msol$, which therefore sets {\it our} halo mass limit.

Galaxies with the highest stellar mass in each group are distinguished
from the rest of the group population under the assumption that they
live at the bottom of the group's potential well ({\it central}
galaxies), whilst the remainder orbit within this potential ({\it
  satellite} galaxies).  Whilst the reality of group dynamics is likely
more complicated \citep[see e.g.][]{Skibba10}, this provides a suitable
comparison sample for our model population for which central and
satellite galaxies are treated differently. Of the 854 galaxies meeting
the stellar mass cut ($\Mstellar \geq 10^{10.5}\Msol$) 810 have
estimated halo masses from \citet{Yang07}, of which 665 are centrals
and 145 are satellites. In any group catalogue, there will be some
misclassification of infalling central galaxies as satellites, or
massive satellite galaxies as centrals. This will only serve to reduce
differences between the satellite and central population in
observations when compared to the models.

Central galaxies can be found in halos down to the sample limit of
$\Mhalo \sim 10^{11.75}\Msol$. A satellite galaxy of the same stellar
mass has at least one more massive galaxy in the halo -- as discussed
in Section~2.8 of \citetalias{Wilman12}, early-type satellites can
reside in halos down to $\Mhalo \sim 10^{12.5}\Msol$.

Table~\ref{table:morphologies} presents the morphological
classifications for our SDSSRC3 sample.

\begin{table*}
\begin{center}
  \caption{A subset of the SDSSRC3 sample morphological
    classifications. The full table can be accessed via the online
    edition of this paper. RA, Dec and redshift are from SDSS-DR4 and
    stellar masses computed using the SDSS photometry and the
    color-based mass-to-light ($M/L$) ratios of \citet{Zibetti09} (see
    text). Note: Morphological classifications are described using the
    RC3 code, see \citet{RC3}. The \citetalias{Wilman12} morphologies
    are the same as in RC3 \textit{except} where we have reclassified
    galaxies (or provided classifications where none existed). If we
    revised an RC3 disc classification (e.g., S0 to spiral), we
    retained any additional morphological information (rings, etc.) as
    in RC3. }
\label{table:morphologies}
\vspace{0.1cm}
\begin{tabular}{ccccccc}
  \hline\hline
  \noalign{\smallskip}
  {\bf Name} & {\bf RA} & {\bf Dec} & {\bf z} & {\bf log$_{10}(\Mstellar/\Msol)$} & {\bf RC3 Morphology} & {\bf \citetalias{Wilman12} Morphology}\\
  \noalign{\smallskip}
  \hline
  \noalign{\smallskip}
  IC25 & 7.800383 & -0.407338 & 0.0194 & 10.52 & .L?.... & .L?....\\ 
  NGC223 & 10.566143 & 0.845477 & 0.0179 & 10.89 & PSBR0*. & PSBR0*.\\ 
  NGC548 & 21.510454 & -1.225612 & 0.0180 & 10.70 & .E+..*. & .E.....\\ 
  UGC1072 & 22.433275 & -1.241424 & 0.0173 & 10.90 & .L..... & .L.....\\ 
  UGC1120 & 23.510277 & -1.075758 & 0.0155 & 10.62 & .SB.2P/ & .SB.2P/\\ 
  NGC1194 & 45.954587 & -1.103726 & 0.0136 & 11.21 & .LA.+*. & .P.....\\ 
  NGC359 & 16.070688 & -0.764911 & 0.0179 & 10.95 & .L..-*. & .E.....\\ 
  NGC364 & 16.170124 & -0.802756 & 0.0171 & 11.14 & RLBS0*. & RLBS0*.\\ 
  UGC1698 & 33.082046 & -0.811519 & 0.0408 & 11.47 & PSBR1*. & PSBR1*.\\ 
  NGC856 & 33.409837 & -0.717291 & 0.0201 & 11.04 & PSAT0*. & PSAT0*.\\ 
  \noalign{\smallskip}
  \hline\hline
\end{tabular}
\end{center}
\end{table*}

Luminosity distances are computed assuming a $\Lambda$CDM cosmology
with $H_0=74.3\kmsMpc$, $\Omega_m=0.226$ and $\Lambda_0=0.774$.

\subsection{Passive Galaxies}\label{sec:passiveobs}

Although \citetalias{Wilman12} paid particular attention to the
relative fractions of S0 galaxies versus those of spirals and
ellipticals, there is no way to identify which model galaxies with
discs currently have spiral structure (the key feature distinguishing
spirals from S0s). Instead, we will concentrate in this paper on 1)
identifying which galaxies are elliptical; and 2) distinguishing which
disc galaxies are active and which are passive in terms of current
star-formation activity.

To define star formation activity, we use the MPA-JHU DR7 calibration
of total specific star formation rates (SSFR, the rate of star
formation per unit stellar mass) for SDSS
galaxies.\footnote{http://www.mpa-garching.mpg.de/SDSS/DR7/} The total
SSFR is computed based on the fiber spectroscopy plus a correction
which attempts to estimate the star formation outside the fiber
aperture.

Emission lines can only be used to calibrate SSFR where there is no
additional source of ionizing radiation.  The MPA-JHU catalogue
includes spectral classification of galaxies, defined using emission
line ratios. This classification -- based on the $\frac{\rm
  [OIII]\lambda5007}{H\beta}$ versus $\frac{\rm
  [NII]\lambda6584}{H\alpha}$ BPT diagram \citep{Baldwin81} -- affects
how the SSFR is estimated. Galaxies are defined in this scheme as {\it
  star-forming} if they lie on the tight locus of normally star-forming
galaxies \citep{Kauffmann03}. Galaxies with a harder radiation field
(higher ratios of $\frac{\rm [OIII]\lambda5007}{H\beta}$ and $\frac{\rm
  [NII]\lambda6584}{H\alpha}$) are further subdivided depending on
whether the line ratios are within the range of \citet{Kewley01}
starburst models. Those which are, are labelled {\it composite} while
galaxies with the hardest radiation fields are labelled {\it AGN}. As
illustrated in Figure~9 of \citetalias{Wilman12}, some of these have
Seyfert nuclei, but most contain Low Ionization Nuclear Emission-line
Regions (LINERS).

Using the fiber spectroscopy, the star formation rate for galaxies
spectroscopically classified as star-forming is estimated from
emission-line modelling. For galaxies with AGN or composite spectra,
and galaxies with no emission lines, the star formation rate is instead
estimated using the D4000$_n$ feature (the strength of the break in the
spectrum at 4000\AA). The strength of this break depends upon the
presence of young ($\lesssim 1~\Gyr$ old) stars which add flux to the
blue part of the spectrum, weakening the break.  This can be compared
to the very young ($\lesssim 20~\Myr$ old) massive stars which are hot
enough to ionize the surrounding gas which then emits light via
recombination lines (such as H$\alpha$). Clearly these two methods
trace star formation on very different timescales.

The MPA-JHU DR7 calibrations include an improved method for aperture
corrections compared to the published DR2 version \citep{Brinchmann04}.
Star formation rates outside the SDSS fibre are estimated by modelling
the observed broad-band colours.

We choose to define galaxies with SSFR $< 10^{-11}$ yr$^{-1}$ ($ <
10^{-2}$ Gyr$^{-1}$, i.e. $\geq 7.5 \times$ below the past averaged star
formation rate at $z=0$) as passive, for the purposes of comparison with
the model galaxies. This level is consistent with definitions of
``passive'' commonly used in the literature
\citep[e.g.][]{Weinmann10,McGee11}.

We now take a moment to consider the distribution of observed galaxies
in derived SSFR, spectral classification, EW[H$\alpha$] and morphology.
The full distribution is complex and an accurate picture of massive
galaxy evolution requires consideration of these and more parameters.
For this purpose we provide a census of this population below. Readers
only concerned with the fraction of galaxies with SSFR $< 10^{-11}$
yr$^{-1}$ can skip to the last paragraph of this section.

\begin{figure*}
  \centerline{\includegraphics[width=\textwidth]{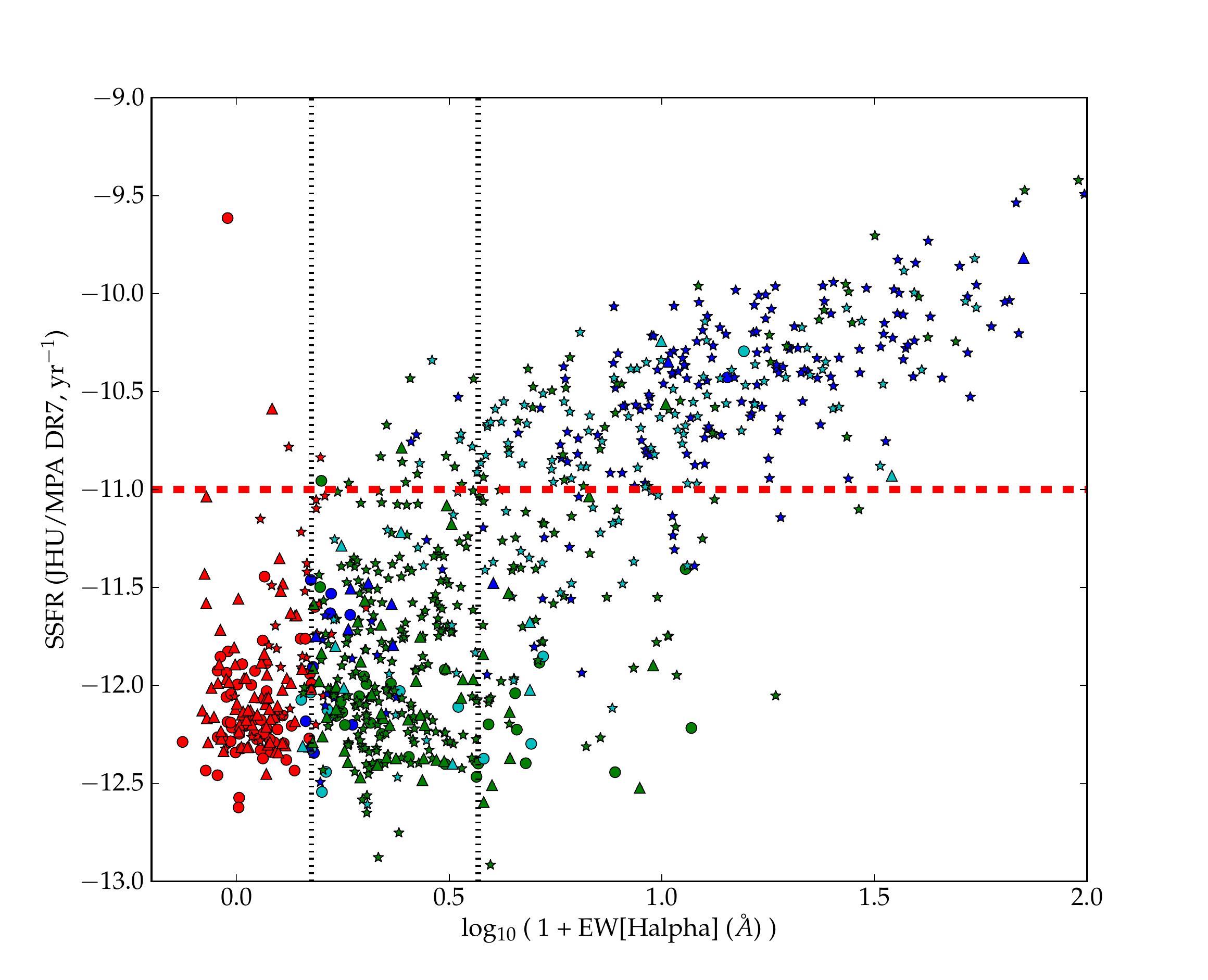}}
  \caption{EW[H$\alpha$] versus MPA-JHU DR7 derived total SSFR for the
    $\Mstellar\geq10^{10.5}\Msol$ SDSSRC3 galaxies, keyed by morphology
    (circles for ellipticals, triangles for S0s, and stars for spirals)
    and spectral classification (blue for star-forming, red for
    H$\alpha$ nondetections at the 2$\sigma$ level, green for AGN, and
    cyan for composite).  We actually plot log$_{10}$(1 +
    EW[H$\alpha$]) with emission defined to be positive to enable the
    data to be log-scaled over the full range for clarity. Our division
    for passive galaxies at SSFR $= 10^{-11}$ yr$^{-1}$ is shown as a
    horizontal red dashed line. We also show the average value of a
    $2\sigma$ detection threshold in H$\alpha$, EW[H$\alpha$] $= 0.5$\AA,
    as well as the value beyond which the H$\alpha$ emission is
    inconsistent with ionization from old stellar populations --
    EW[$H\alpha$] $= 3$\AA, \citep{CidFernandes11} (vertical black dotted
    lines). 308 of 406 galaxies with a $> 2\sigma$ detection of the
    H$\alpha$ emission line and SSFR $< 10^{-11}$ yr$^{-1}$ are AGN and 297
    of this 406 are spirals (there are also 63 S0s and 46
    ellipticals).}
  \label{fig:hassfrobs}
\end{figure*}

Our method of defining ``passive'' versus star-forming (in terms of
SSFR) should not be assumed to translate into a simple case of
``passive = no emission lines'', and one should not assume that spirals
are automatically ``star-forming''.  Figure~\ref{fig:hassfrobs} shows
SSFR versus H$\alpha$ equivalent width (EW[H$\alpha$]) for the SDSSRC3
galaxies, which are keyed by their morphology (circles for ellipticals,
triangles for S0s, and stars for spirals) and their MPA-JHU DR7 spectra
classifications (blue for star-forming, red for H$\alpha$ nondetections
at the 2$\sigma$ level\footnote{Errors are scaled up by a factor 2.473
  as calibrated by \citet{Brinchmann04} using repeated measurements of
  the same galaxy.}, green for AGN, and cyan for composite). The
EW[H$\alpha$] measurements come from the MPA-JHU DR7 fiber calibration
which corrects the emission flux for underlying stellar absorption. We
set emission to be positive, and apply a minor correction of $-0.3$\AA\
to put the peak for passive galaxies at $0$\AA.

What Figure~\ref{fig:hassfrobs} demonstrates is that while almost all
ellipticals and S0s are ``passive'' in terms of our SSFR threshold, so
are many emission-line spiral galaxies. In fact, $54\%$ of the SDSSRC3
spiral galaxies are passive.

The red symbols (159 galaxies lacking significant H$\alpha$ emission)
are mostly ellipticals (59) and S0s (69) but also some spirals (31).
Just four have SSFR $> 10^{-11}$ yr$^{-1}$.

Galaxies with SSFR $> 10^{-11}$ yr$^{-1}$ mostly have significant detections
of the H$\alpha$ emission line (289 out of 293) and are almost all
spiral galaxies (280 of the 293, $96\%$), spectrally classified as
either star-forming, composite or AGN galaxies. There is a strong, if
rather broad correlation between SSFR and EW[H$\alpha$].

Fully $72\%$ of the passive galaxies have detectable H$\alpha$ emission
(406 galaxies); of these, $76\%$ (308) are classified as AGN.  Ionizing
radiation from old stellar populations may be enough to explain
H$\alpha$ emission in galaxies with EW[H$\alpha$]$\lesssim3$\AA
\citep[][vertical dotted line at 2.7\AA]{CidFernandes11}\footnote{The
  line is plotted at 2.7\AA\ to account for our correction of
  EW[H$\alpha$] by $-0.3$\AA.} and AGN-like line ratios. This may
account for up to $80\%$ of ``AGN'' in passive galaxies with detectable
H$\alpha$ emission.

$73\%$ (297) of passive galaxies with detectable H$\alpha$ emission
have spiral morphology. However, many passive elliptical (46) and S0
(63) galaxies also have significant detections of the H$\alpha$
emission line.  These are typically at low SSFR ($< 10^{-11.5}$ yr$^{-1}$)
and EW[H$\alpha$]$\lesssim 3$\AA (consistent with ionizing radiation
from old stellar populations).

Spiral galaxies contribute $87\%$ of the passive galaxies just below
our division: i.e. with $10^{-11.5}$ yr$^{-1} <$ SSFR $< 10^{-11}$ yr$^{-1}$.
These spirals contribute $31\%$ of all passive spiral galaxies
demonstrating that the fraction of passive spirals is very sensitive to
the definition of ``passive''.  In other words, quenching of galaxies
doesn't necessarily mean that star formation rates go to zero: some
galaxies continue forming stars at lower rates.  The emission line
ratios of such galaxies lead to them being predominantly classified as
either AGN or composite.

\begin{figure*}
  \centerline{\includegraphics[width=\textwidth]{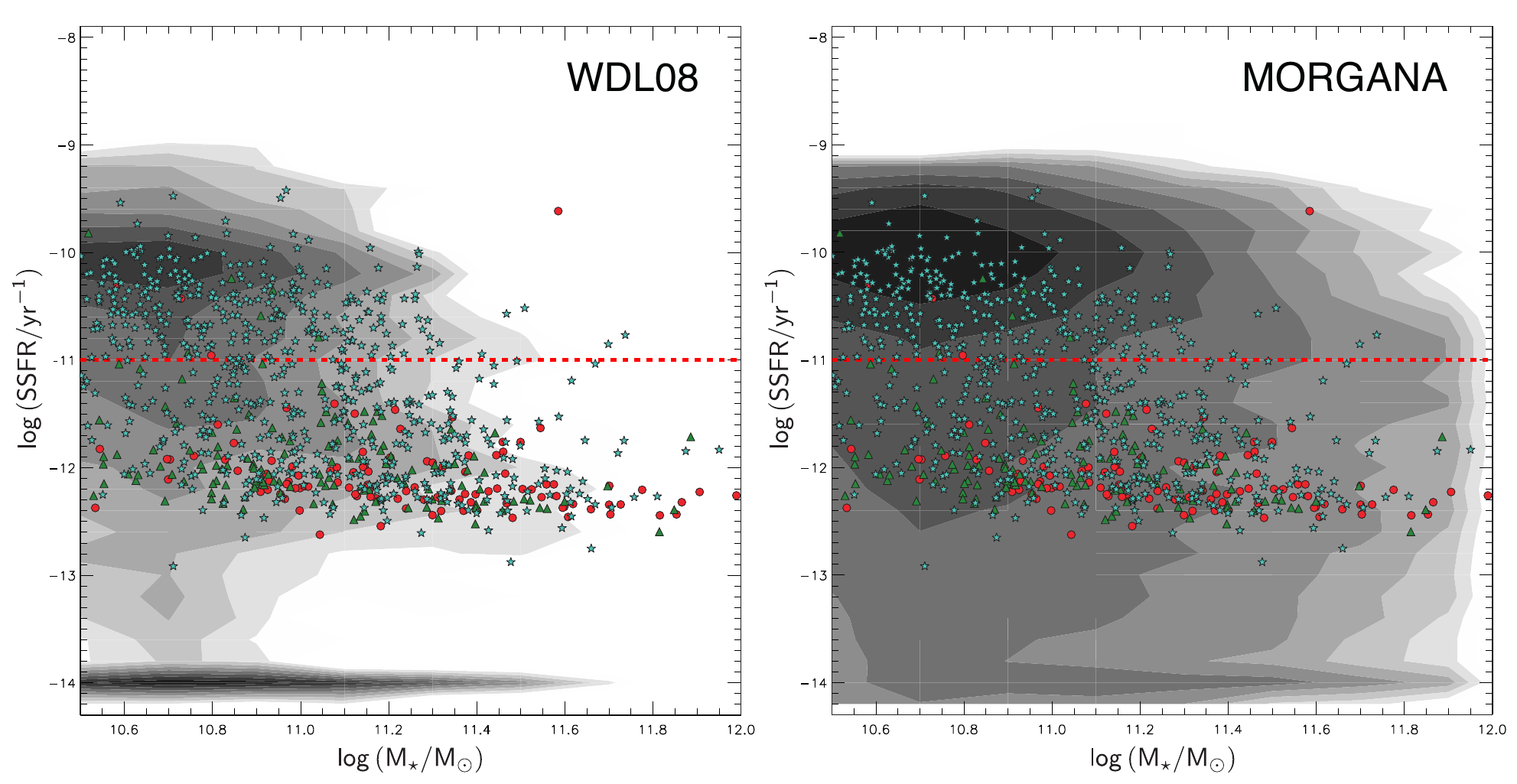}}
  \caption{The distribution of galaxies in specific star formation rate
    of galaxies versus their stellar mass in the pure mergers
    implementation of the \citetalias{WDL08} model (left) and \morgana\ model
    (right). The contour levels are log-spaced, and we set all model
    galaxies with SSFR less than a limiting value of $10^{-14}$ yr$^{-1}$
    equal to that value.  Overplotted are all galaxies from the SDSSRC3
    sample keyed by morphology (red circles for ellipticals; green
    triangles are S0s and cyan stars are spiral galaxies). }
            \label{fig:ssfrms}
\end{figure*}

Figure~\ref{fig:ssfrms} shows the distribution of SSFR of all
$\Mstellar > 10^{10.5}\Msol$ galaxies versus stellar mass ($\Mstellar$)
at $z=0$ for each model (contours) with the pure merger implementation
for bulge growth, with the SDSSRC3 sample overplotted (symbols).  The
distribution of model galaxies in this plane does not change
significantly with the particular bulge growth implementation used,
producing a star-forming population for which SSFR is roughly
independent of stellar mass, with a typical SSFR $\sim 10^{-10}$ yr$^{-1}
= 0.1$ Gyr$^{-1}$. This corresponds to a mass-doubling rate of around the
Hubble time and is $\sim 0.5$ dex above the SSFR for the average star
forming SDSSRC3 galaxy. Both models and observations exhibit a drop in
number density at around our division of SSFR $= 10^{-11}$ yr$^{-1}$. Below
this cut, the SSFR of SDSSRC3 passive galaxies are limited to a minimum
SSFR $\sim 10^{-12.5}$--$10^{-12}$ yr$^{-1}$, representing a limit to the
template fitting procedure of \citet{Brinchmann04}. For clarity of
presentation, we display all model galaxies with SSFR $< 10^{-14}$ yr$^{-1}$
as having SSFR $= 10^{-14}$ yr$^{-1}$; this includes galaxies in
\citetalias{WDL08} which have \textit{zero} SSFR ($39\%$ of
\citetalias{WDL08} galaxies).  In contrast, all \morgana\ galaxies are
forming some stars, with very few at or below our limiting
SSFR $= 10^{-14}$ yr$^{-1}$ value.

\section{Bulge to Total ratios}\label{sec:btmass}


Detailed photometric decompositions with the resolution necessary to
distinguish and accuractely characterize discs and bulges -- and
especially to distinguish classical (pressure-supported, sersic
parameter $n\sim 4$) from pseudo (rotating, flat, $n\lesssim2$) -type
bulges -- are not available for large samples of galaxies. One high
quality, volume-limited dataset is provided by \citet{FD11}.  They
provide decompositions at $3.6\mu$m of all relatively massive,
non-edge-on galaxies within the local $11\Mpc$ volume, including
classifications of bulges as either ``classical'' or ``pseudo''. The
mass to light ratio varies very little with the stellar population at
$3.6\mu$m, and so $\bt$ at $3.6\mu$m is a reasonable proxy for $\bt$ in
stellar mass.

Elliptical galaxies are assumed to have $\bt=1$. It is difficult to
discern disc components in galaxies classified as ellipticals: if any
disc exists, it will typically be embedded in the bulge, and comprise
only a few percent of the galaxy's stars \citep[e.g.][]{Scorza98}.
Disc galaxies (S0s and spirals) have a much higher mass fraction in
their discs, with $\bt$ in the range 0--0.5 for spirals (with
increasing numbers to low $\bt$ ) and 0--0.7 for S0s. This is
confirmed with larger samples of nearby disk galaxies
\citep[e.g.][]{Weinzirl09,Laurikainen10}.

\begin{figure*}
  \centerline{\includegraphics[width=\textwidth]{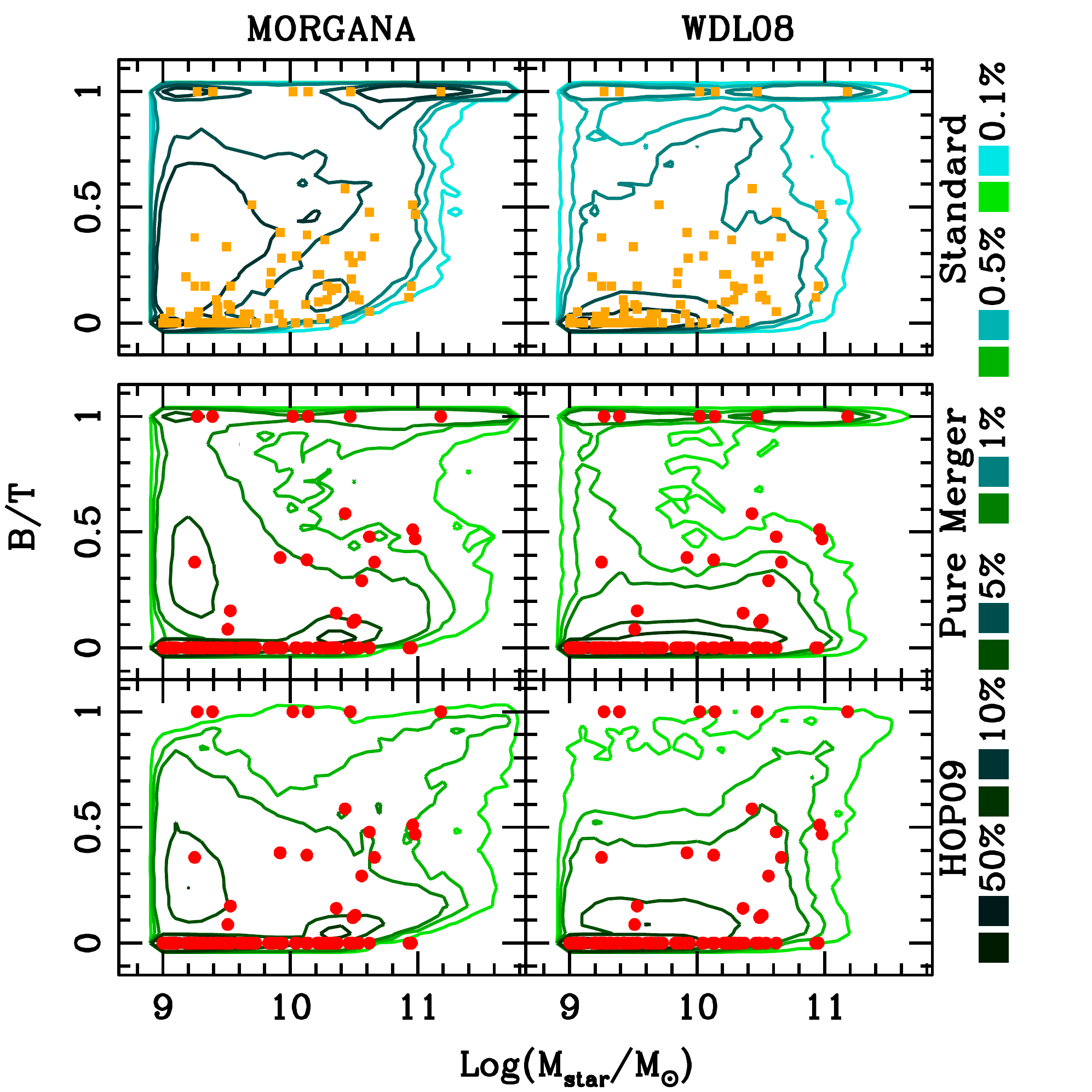}}
  \caption{Contour plot showing the distribution in bulge to total
    ratio ($\bt$) vs stellar mass ($\Mstellar$) for each model
    implementation.  Overplotted points are real galaxies from the
    \citet{FD11} 11~Mpc sample. The top panels compare the models with
    standard bulge growth implementations (includes disc instabilities,
    turquoise contours) to the total (pseudo$+$classical) bulge to
    total ratio of observed galaxies (orange points), while the middle
    and lower panels compare the models with pure merger and \hop\
    bulge formation implementations (green contours) to the observed
    classical bulge to total ratio (red points).}
  \label{fig:fd11comp}
\end{figure*}

Figure~\ref{fig:fd11comp} compares the $\bt$ stellar mass ratios
($\bt$) of model galaxies with the equivalent $3.6\mu$m luminosity
ratios of \citet{FD11}. Six panels are presented, one for each bulge
formation implementation applied to each model. Contours describe the
full distribution of model galaxies in $\bt$ vs stellar mass
($\Mstellar$).

Whereas classical bulges are thought to form via galaxy mergers,
pseudo-bulges likely result from disc instabilities \citep{Kormendy04}.
We can therefore link our different channels of bulge formation to the
observed category of bulge. Comparing to our pure merger and \hop\
implementations (middle and lower panels), we plot the {\it classical}
bulge to total ratio (i.e. $\bt>0$ only for bulges classified by
\citet{FD11} as ``classical''). In the standard model (upper panels),
bulges form via both mergers and disc instabilities, and so we compare
to the total ({\it pseudo$+$classical}) bulge to total ratio from
\citet{FD11}.

The pure mergers implementations predict a bimodal distribution of
$\bt$ for $\Mstellar \gtrsim 10^{10}\Msol$. It peaks at $\bt\sim1$
(ellipticals) and then there is a gap at $0.55\lesssim\bt\lesssim0.95$.
For $\bt<0.55$, the fraction increases with decreasing $\bt$ such that
there is a significant population of almost ``bulgeless'' galaxies.  In
\citet{Fontanot11} we find a reasonable match for pure merger models to
the fraction of observed classical-bulgeless galaxies as a function of
stellar mass, comparing with data from \citet{FD11} and from
\citet{Kormendy10}.  Now we also see that the full distribution of
$\bt$ is well matched. We note that the {\it fraction} of bulgeless
galaxies is well matched at all masses, even at
$\Mstellar<10^{10}\Msol$ \citep{Fontanot11} despite the low number of
observed galaxies with significant bulges at this mass.

The gap in the distribution of $\bt$ for galaxies in the pure mergers
implementations results from the total destruction of discs in major
mergers. The \hop\ implementation allows a fraction of stellar and gas
discs to survive even major mergers. This results in fewer $\bt=1$
galaxies, with major merger remnants overpopulating the
$0.55\lesssim\bt\lesssim0.95$ region, so that the $\bt$ distribution is
more continuous. While we cannot rule out the existence of minor
embedded discs in some elliptical galaxies, we can say that these {\it
  residual discs} are not those in galaxies with $\bt \lesssim 0.7$, as
found in local spirals and S0s \citep{Laurikainen10}.

To compare with our observed sample, {\it we define model elliptical
  galaxies as those with $\bt \geq 0.7$.}  This is sufficient to
distinguish a typical observed disc galaxy from an elliptical.  Pure
merger and \hop\ implementations provide similar elliptical fractions
with this cut, and so we simplify our analysis from here onwards by
considering only the pure mergers implementation. A cut at (e.g.)
$\bt=0.9$ (as in \citetalias{deLucia11}) results in fewer ``ellipticals'' in the \hop\
implementations than with the pure mergers implementations.

Despite problems modelling disc instabilities, the distribution of
$\bt$ with the standard implementations are reasonably well matched to
the classical plus pseudo bulge fractions in Figure~\ref{fig:fd11comp}
(upper panels), although both models produce massive bulges via disc
instabilities (including $\bt\sim1$ galaxies within \morgana) which is
inconsistent with the low mass of most observed pseudo-bulges.

We prefer to focus on the better constrained merger channel for bulge
growth in the next Sections. However, for consistency with the
literature, we also include plots showing the behaviour of the standard
models in Appendix~\ref{sec:stdmodels}.

\section{Morphological Fractions vs Stellar and Halo Mass}\label{sec:results}

In this section, we examine the fraction of galaxies of different
morphological type.  For reference, we publish the total integrated
fractions in Table~\ref{table:totalfractions}. SDSSRC3 types include
visually classified ellipticals, and visually classified S0s$+$spirals
(disc galaxies) defined to be passive or star-forming.  Model fractions
are computed with each bulge growth implementation, and include
ellipticals ($\bt\geq0.7$) and disc galaxies ($\bt<0.7$) defined to be
passive or star-forming. The division between passive and star-forming
is made at $\ssfr = 10^{-11} \yr^{-1}$ (Section~\ref{sec:passiveobs}).

\begin{table*}
\begin{center}
\caption{Total fraction of galaxies of different type, integrated down to $\Mstellar=10^{10.5}\Msol$.}
\label{table:totalfractions}
\vspace{0.1cm}
\begin{tabular}{cccc}
\hline\hline
\noalign{\smallskip}
{\bf Observations} & f(E) & f(passive disc) & f(star-forming disc)\\
\noalign{\smallskip}
\hline
\noalign{\smallskip}
SDSSRC3 & $0.08 \pm 0.01$ & $0.58 \pm 0.02$ & $0.34  \pm 0.02$\\
\noalign{\smallskip}
\hline
\noalign{\smallskip}
{\bf Model$+$implementation} & f(E) & f(passive disc) & f(star-forming disc)\\
\noalign{\smallskip}
\hline
\noalign{\smallskip}

{\bf \citetalias{WDL08}} & & & \\
Standard & $0.308 \pm 0.005$ & $0.277 \pm 0.004$ & $0.415 \pm 0.005$\\ 
Pure merger & $0.204 \pm 0.004$ & $0.362 \pm 0.005$ & $0.434 \pm 0.005$\\
\hop\ & $0.187 \pm 0.004$ & $0.415 \pm 0.005$ & $0.399 \pm 0.005$\\
\morganabf\  & & & \\
Standard & $0.642 \pm 0.003$ & $0.102 \pm 0.002$ & $0.257 \pm 0.003$\\
Pure merger & $0.334 \pm 0.003$ & $0.154 \pm 0.002$ & $0.512 \pm 0.003$\\
\hop\ & $0.227 \pm 0.003$ & $0.179 \pm 0.003$ & $0.595 \pm 0.003$\\
\morganabf\ {\bf longer $\tausat$} & & & \\
Standard & $0.641 \pm 0.003$ & $0.134 \pm 0.002$ & $0.226 \pm 0.003$\\
Pure merger & $0.297 \pm 0.003$ & $0.216 \pm 0.003$ & $0.487 \pm 0.003$\\
\hop\  & $0.147 \pm 0.002$ & $0.264 \pm 0.003$ & $0.589 \pm 0.003$\\
\noalign{\smallskip}
\hline\hline
\end{tabular}
\end{center}
\end{table*}

We shall now examine how morphological fractions depend upon stellar
and halo mass, separately for central and satellite galaxies.  SDSSRC3
data is compared with the pure merger implementation for bulge growth
applied to both \citetalias{WDL08} and \morgana\ models, and to the
\morgana\ model with longer satellite survival times. Results for the
standard bulge growth implementations (including disc instabilities)
are presented in Appendix~\ref{sec:stdmodels}.

\subsection{Elliptical Fraction}\label{sec:ellipticals}

As described in Section~\ref{sec:btmass}, model ellipticals are defined
to have $\bt\geq0.7$. We now compare their abundance with that of
visually classified SDSSRC3 ellipticals.

\begin{figure*}
  \centerline{\includegraphics[width=0.49\textwidth]{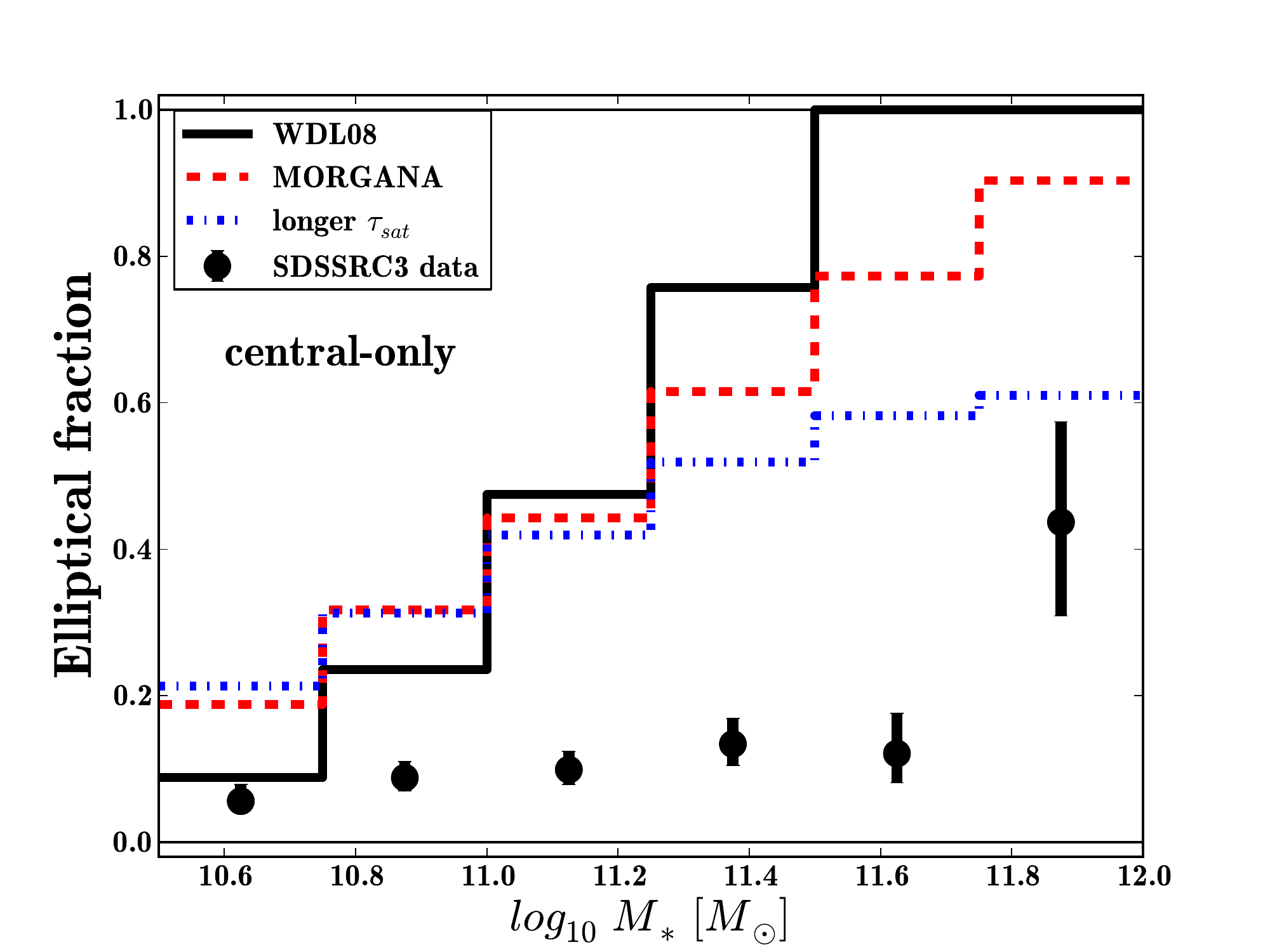}
            \includegraphics[width=0.49\textwidth]{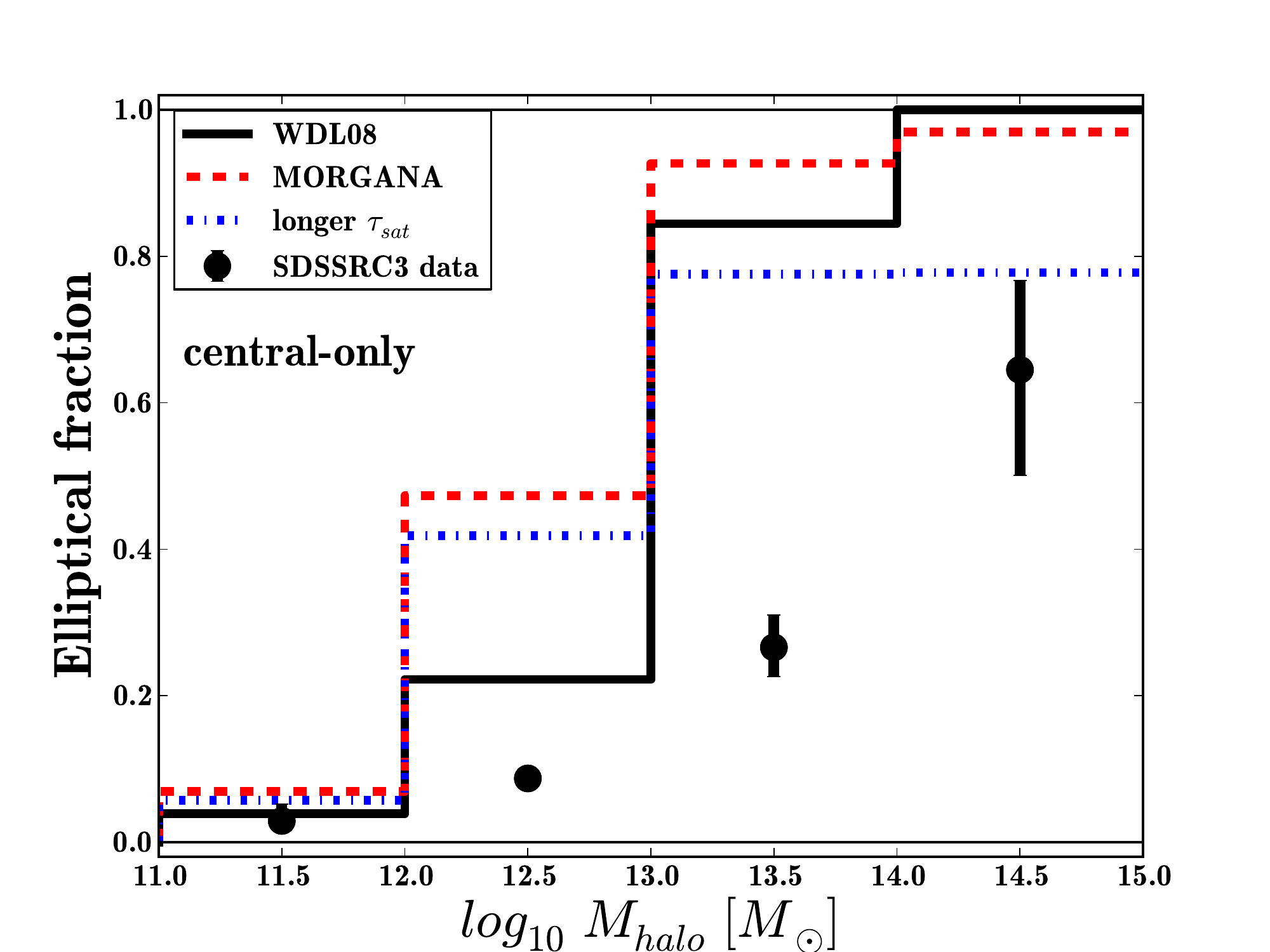}
            }
  \centerline{\includegraphics[width=0.49\textwidth]{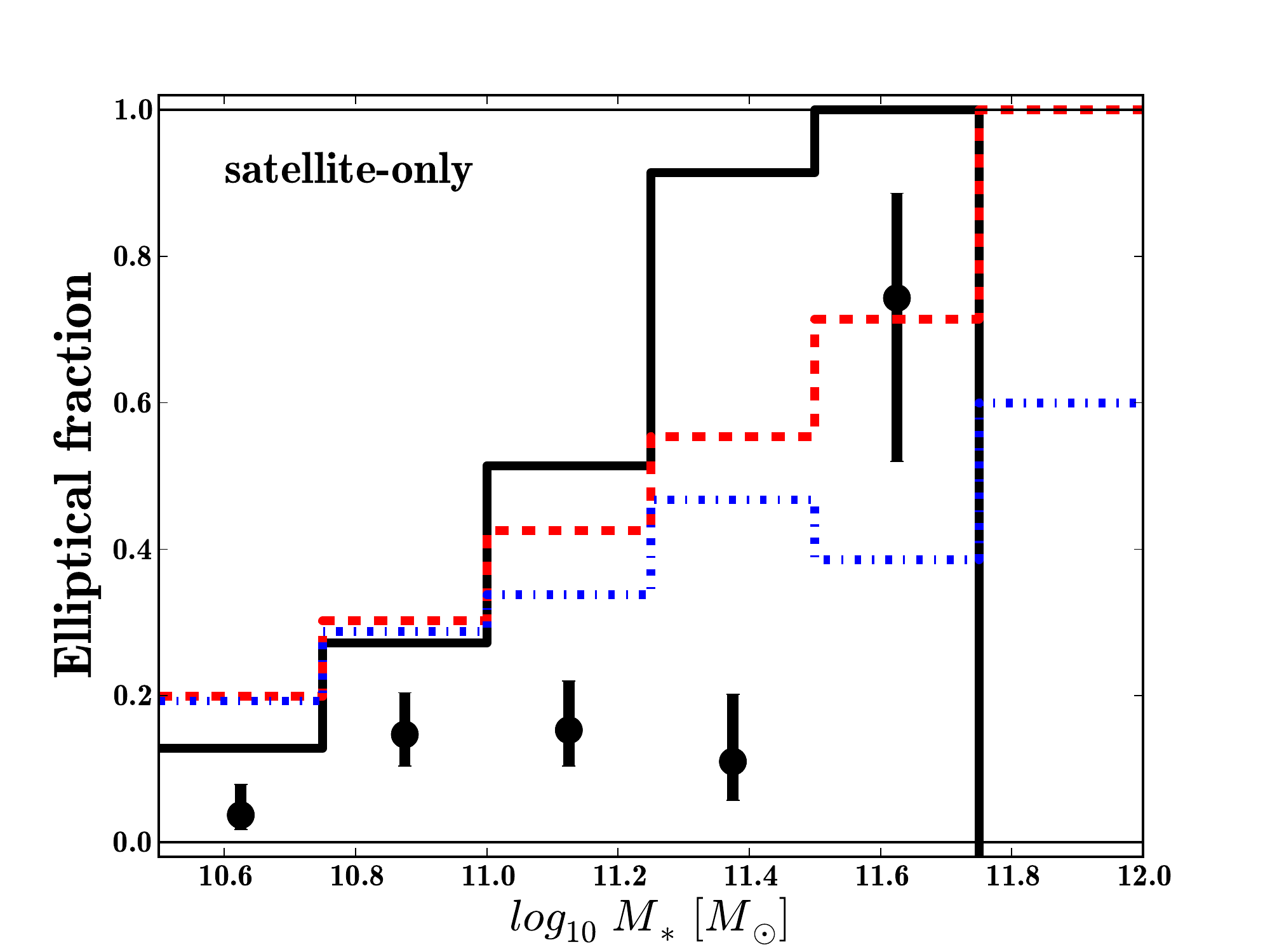}
              \includegraphics[width=0.49\textwidth]{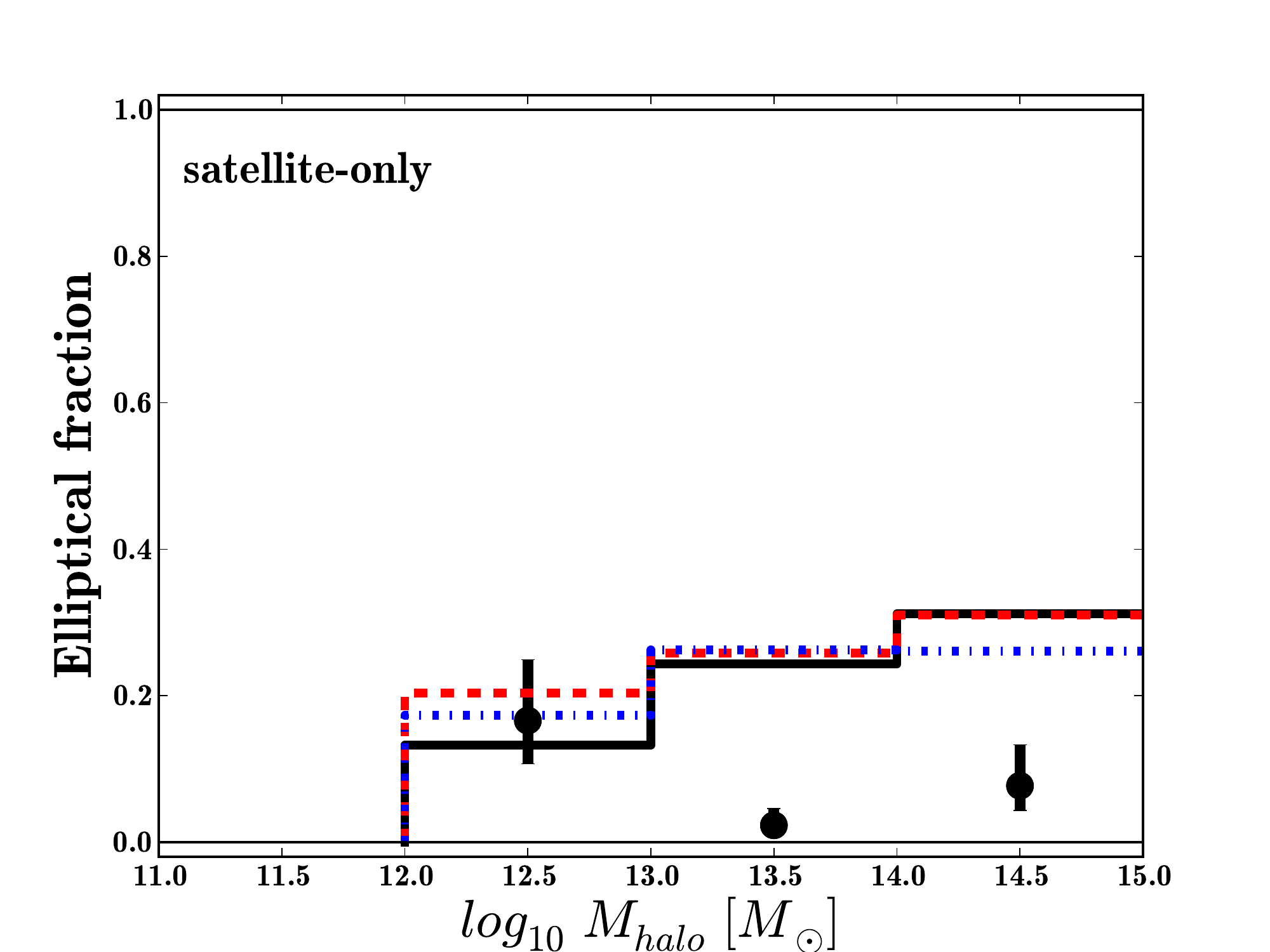}
            }

            \caption{Elliptical galaxy fraction (black points, SDSSRC3
              sample) for $\Mstellar \geq 10^{10.5}\Msol$ galaxies as a
              function of stellar mass ($\Mstellar$, left panels) and
              halo mass ($\Mhalo$, right panels). In the top row we
              only consider central galaxies and in the bottom row we
              only consider satellite galaxies. This is compared with
              the fraction of model elliptical galaxies ($\bt\geq0.7$)
              in the pure mergers implementations of \citetalias{WDL08}
              (solid black line) and \morgana\ (dashed red line) models
              and the \morgana\ model with longer satellite survival
              times (dot-dashed blue line).  Errors are 1$-\sigma$
              binomial errors based on the \citet{Wilson27}
              approximation.}
  \label{fig:fEz0}
\end{figure*}

Figure~\ref{fig:fEz0} shows how the fraction of $\Mstellar \geq
10^{10.5}\Msol$ galaxies which have elliptical morphology depends upon
stellar mass ($\Mstellar$, left panels) and halo mass ($\Mhalo$, right
panels), independently for central galaxies only (upper panels) or
satellite galaxies only (lower panels).  Each figure shows the observed
fraction of SDSSRC3 galaxies visually classified as ellipticals --
black points with 1$-\sigma$ binomial errors based on the
\citet{Wilson27} approximation\footnote{As described by
  \citetalias{Wilman12}, we estimate the uncertainties by first
  rescaling all (weighted) counts so that the total counts are equal to
  the original (unweighted) total counts in a given bin, and then
  computing the Wilson confidence limits using the rescaled counts.} --
to be constrasted with the models. Model elliptical fractions are
presented for the pure mergers implementation for \citetalias{WDL08}
(solid black line) and \morgana\ (dashed red line) models and the
\morgana\ model with longer satellite survival times (dot-dashed blue
line).  We do not show statistical errors on model fractions to improve
clarity: these are much smaller than those for observed fractions, or
differences between implementations.

Figure~\ref{fig:fEz0} shows that the fraction of model ellipticals
increases with stellar mass for both central and satellite galaxies for
all models. This is also true of the observed elliptical fraction,
although this fraction remains low except in the highest mass bin.
Trends with halo mass are also qualitatively comparible: the fraction
of elliptical galaxies in both models and observations increases with
halo mass for central galaxies, but remains low for satellites.

The stellar mass and bulge fractions of central galaxies grow with
their haloes as shown in \citetalias{deLucia11}: halos merge, leading
eventually to the merger of their central galaxies, and thence to the
formation of bulges and (particularly in the case of major mergers)
elliptical galaxies.  Since the probability that a galaxy has acquired
an elliptical morphology derives from its own growth history, it
correlates with its stellar mass in all cases, but with the parent halo
mass only for a central galaxy.

Despite this success, both Figure~\ref{fig:fEz0} and
Table~\ref{table:totalfractions} make it clear that the fraction of
elliptical galaxies produced by the models is significantly higher than
the observed fraction across our range of $\Mstellar$ and $\Mhalo$.
{\it Ellipticals are overproduced by the models.}  This is even more
true if ellipticals can also be formed via disc instabilities (see
Appendix~\ref{sec:stdmodels}). In fact, model elliptical galaxies are
formed fairly ubiquitously at the centre of $\Mhalo\gtrsim10^{13}\Msol$
halos merely as a consequence of their hierarchical growth and
subsequent merger history, in direct conflict with observations. This
poses a serious challenge for semi-analytic models which we shall try
to address in Section~\ref{sec:interpretation}.

\subsection{Star-Forming and Passive Disc Galaxy Fraction}\label{sec:pdisc}

We now turn to disc galaxies. Model disc galaxies ($\bt<0.7$) are
compared to the observed spiral$+$S0 population. We separately compare
the abundance of star-forming, and passive disc galaxies, divided at
$\ssfr = 10^{-11} \yr^{-1}$(see Section~\ref{sec:passiveobs}).

\begin{figure*}
  \centerline{\includegraphics[width=0.49\textwidth]{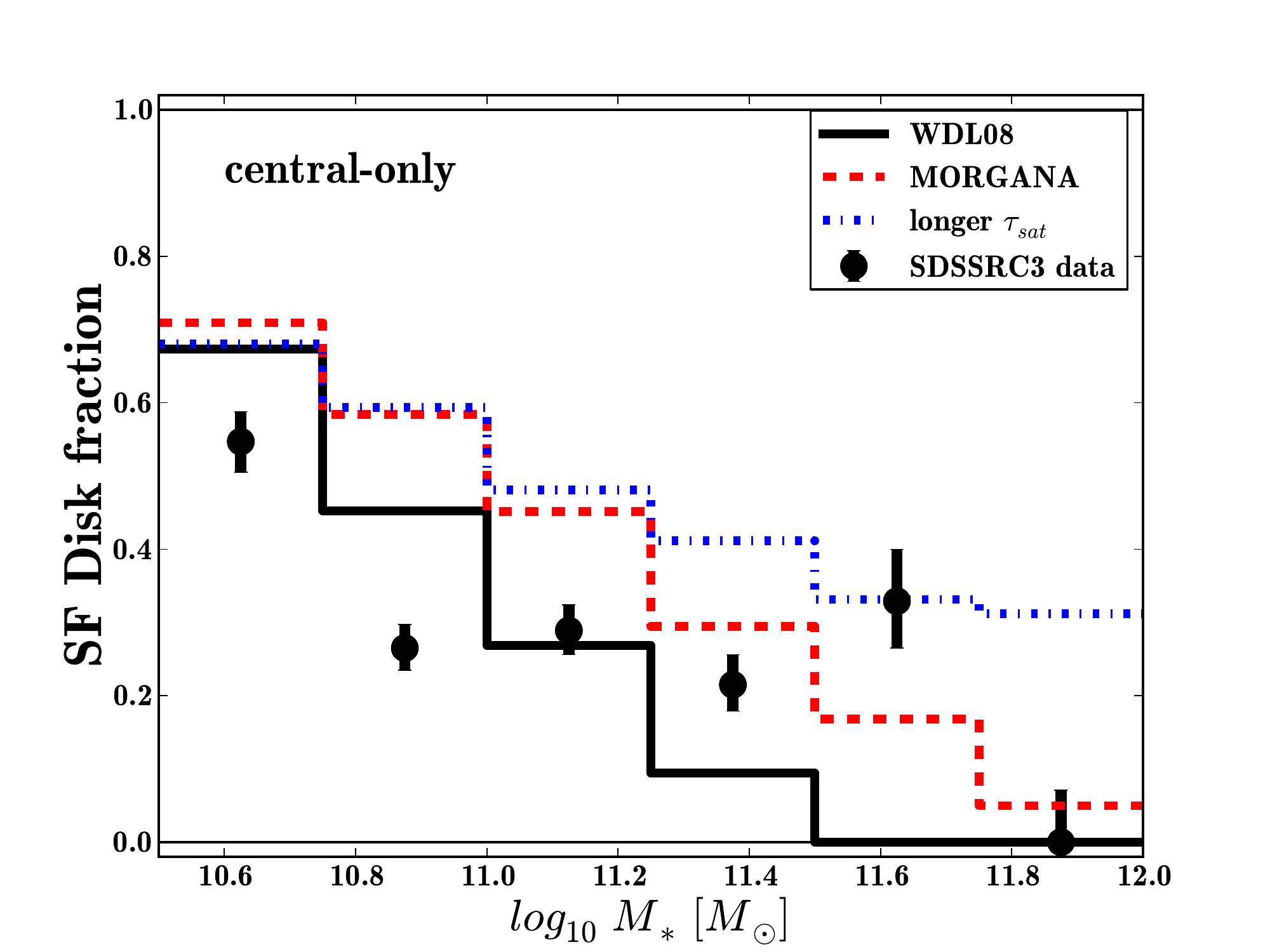}
              \includegraphics[width=0.49\textwidth]{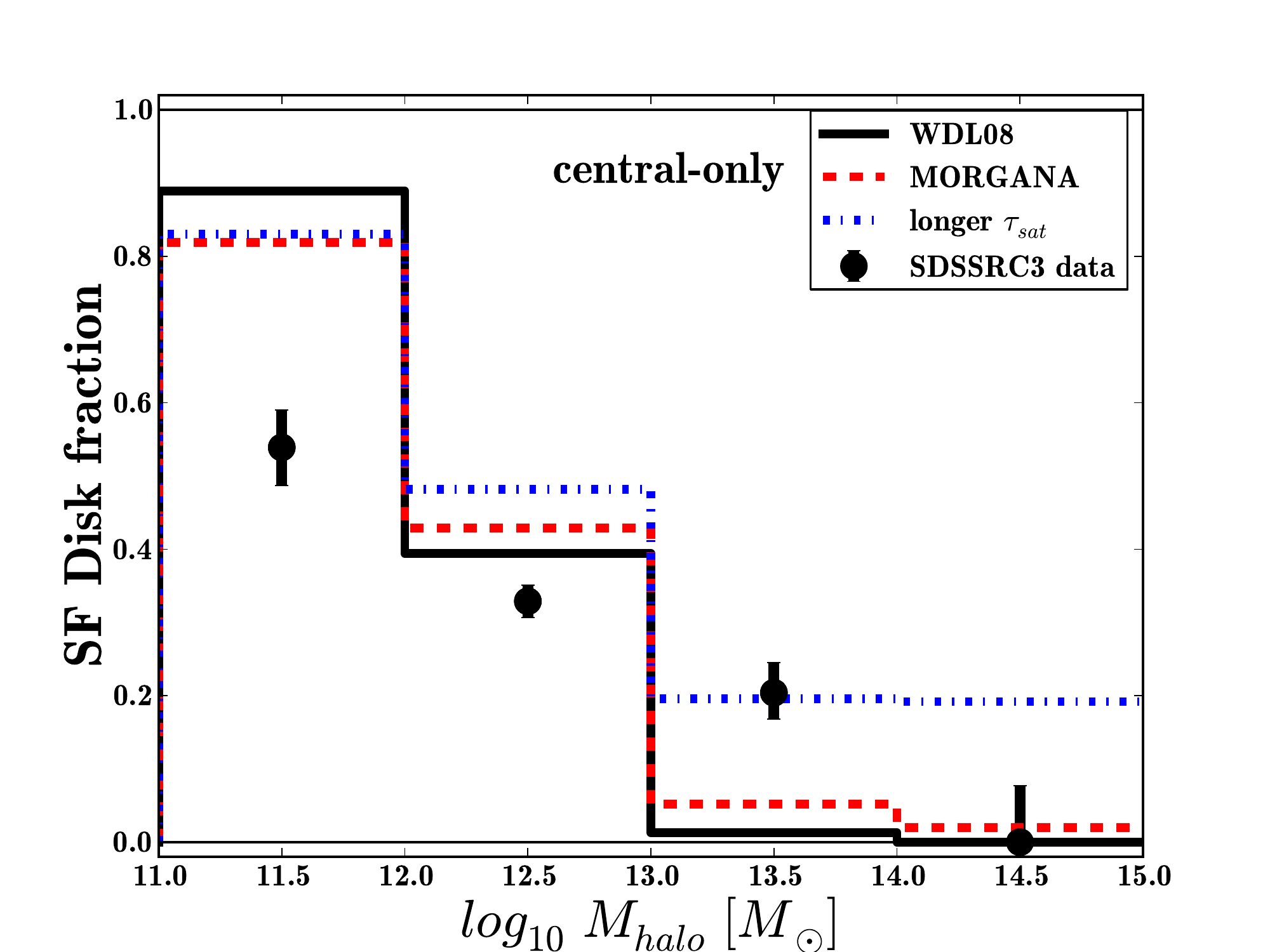}
            }
  \centerline{\includegraphics[width=0.49\textwidth]{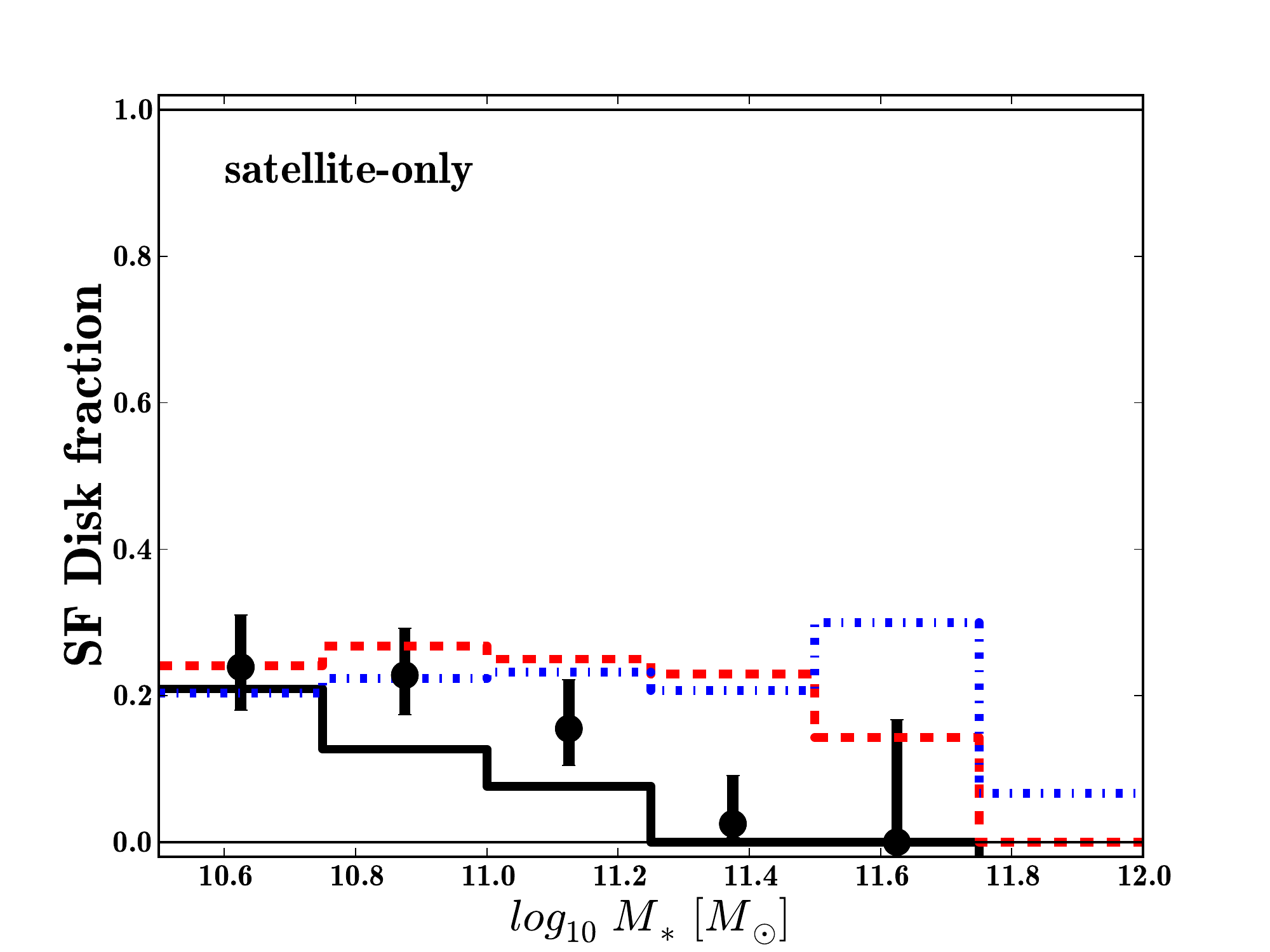}
              \includegraphics[width=0.49\textwidth]{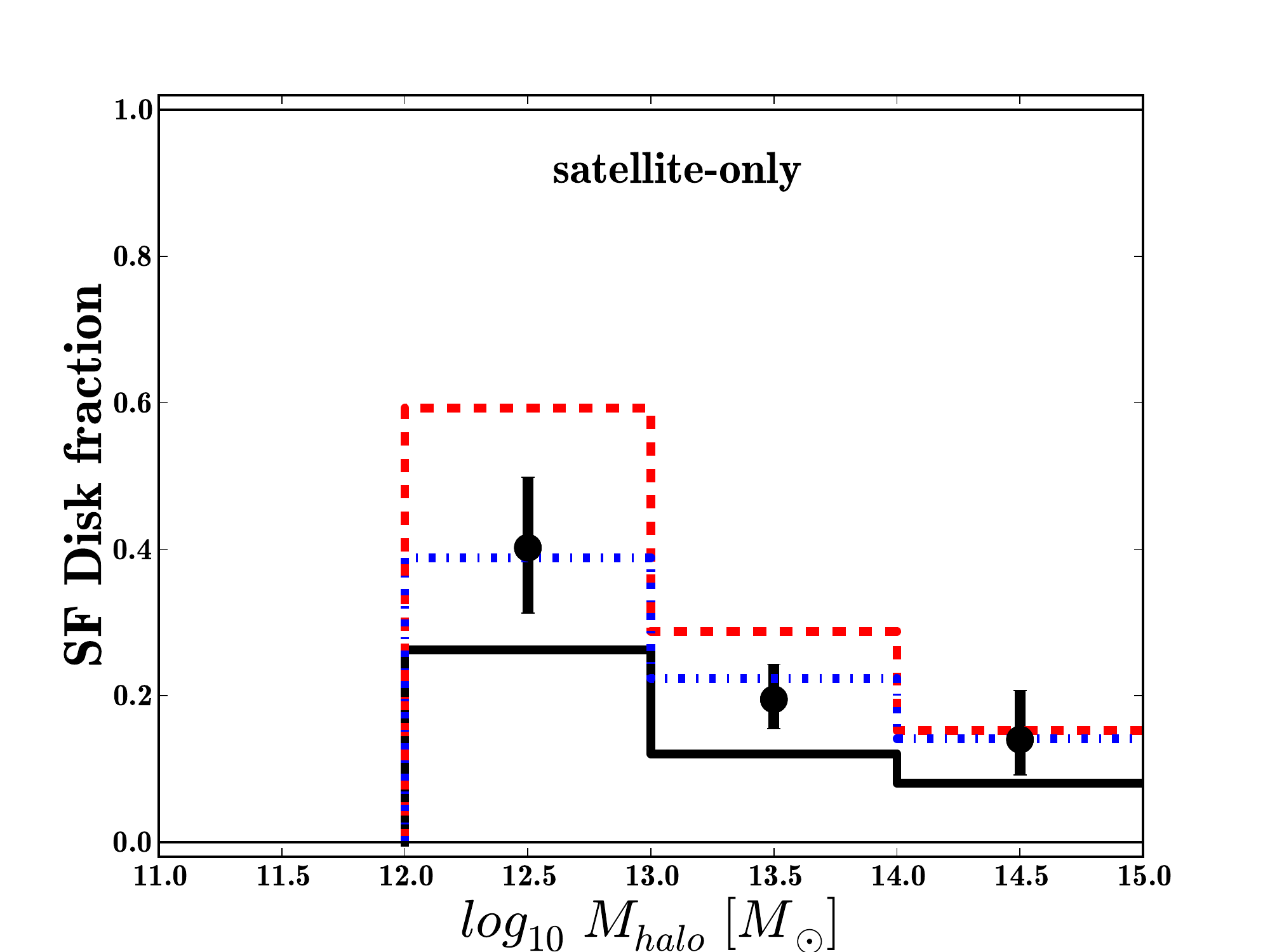}
            }
            \caption{As Figure~\ref{fig:fEz0} except this time we
              present the fraction of star-forming (SSFR $\geq
              10^{-11} \yr^{-1}$) disc galaxies divided into central and
              satellite populations as a function of stellar and halo
              mass and contrasting observations with the pure mergers
              model implementations (see key).}
  \label{fig:fDsfz0}
\end{figure*}

Figure~\ref{fig:fDsfz0} examines the fraction of star-forming disc
galaxies, with the same format as Figure~\ref{fig:fEz0}. This fraction
declines with both stellar and halo mass for both centrals and
satellites in a way which is qualitatively well matched to the data
(except possibly for centrals versus $\Mstellar$).  The only clear
discrepancy is that all models overproduce star-forming disc galaxies
at the centre of $\Mhalo<10^{12}\Msol$ halos. Altogether, \morgana\
produces more star-forming disc galaxies than \citetalias{WDL08},
especially with the longer $\tausat$ and at high mass. However,
comparison to data suggests no clear preference between the models with
current statistics.

\begin{figure*}
  \centerline{\includegraphics[width=0.49\textwidth]{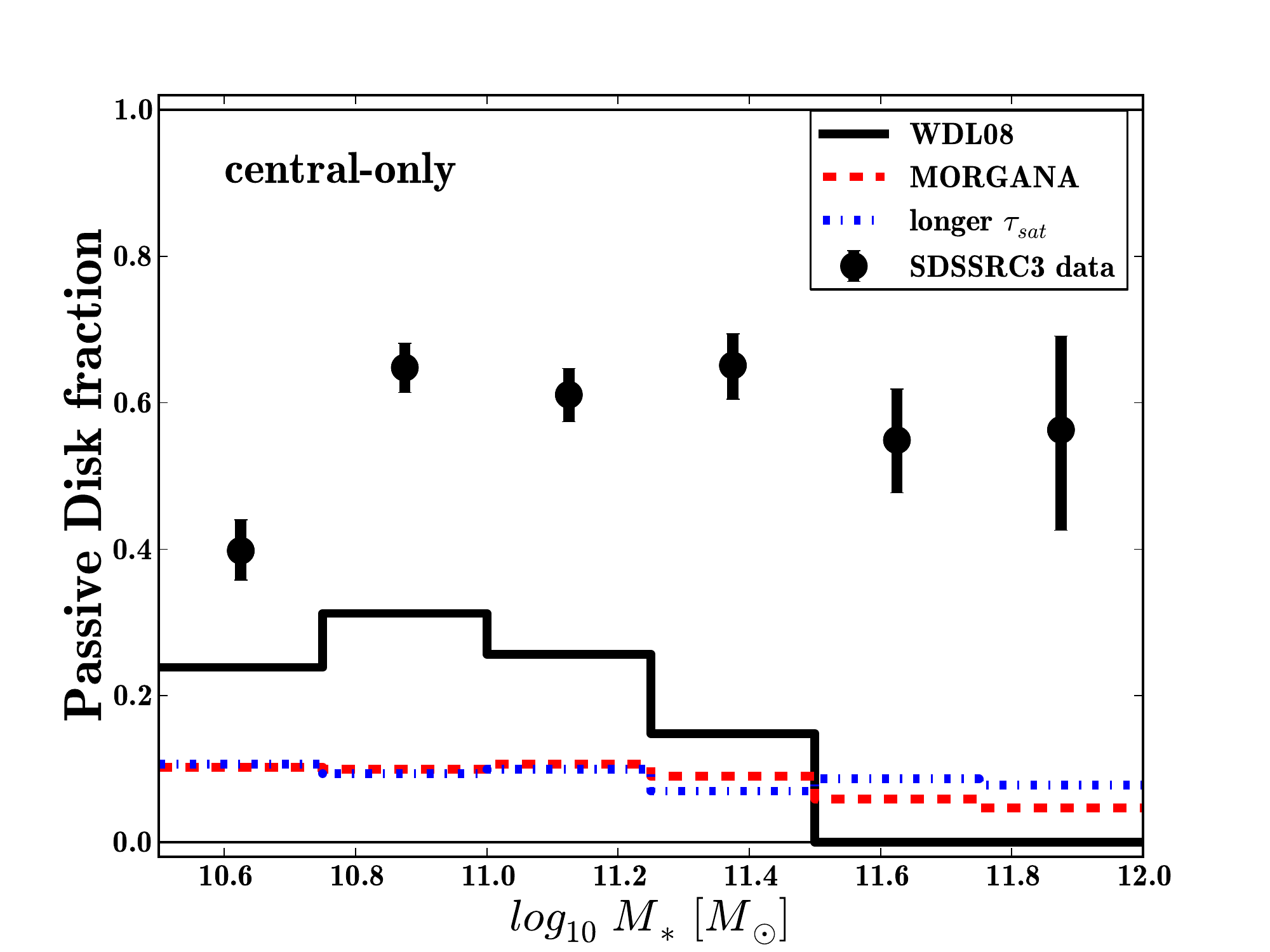}
              \includegraphics[width=0.49\textwidth]{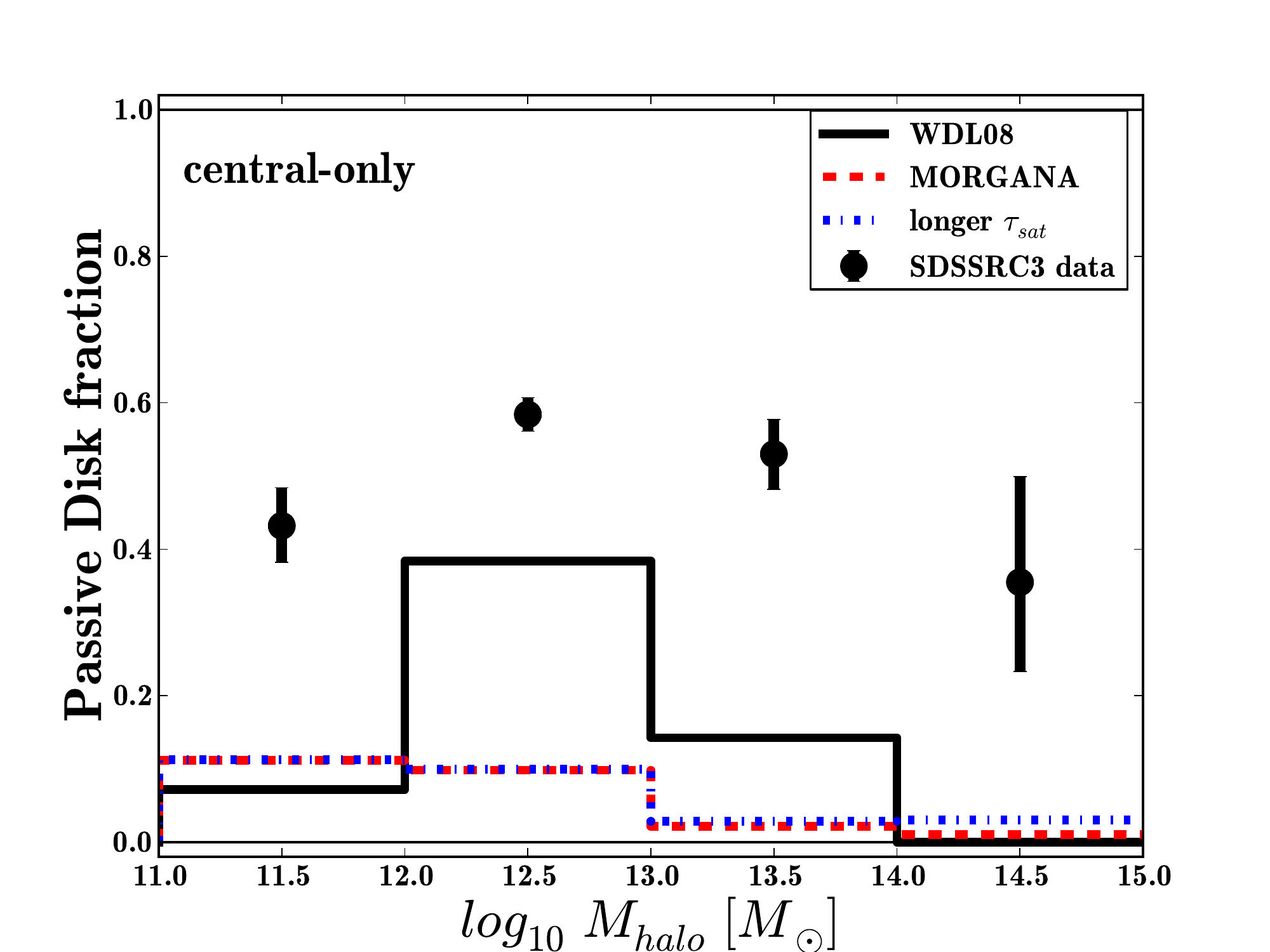}
            }
  \centerline{\includegraphics[width=0.49\textwidth]{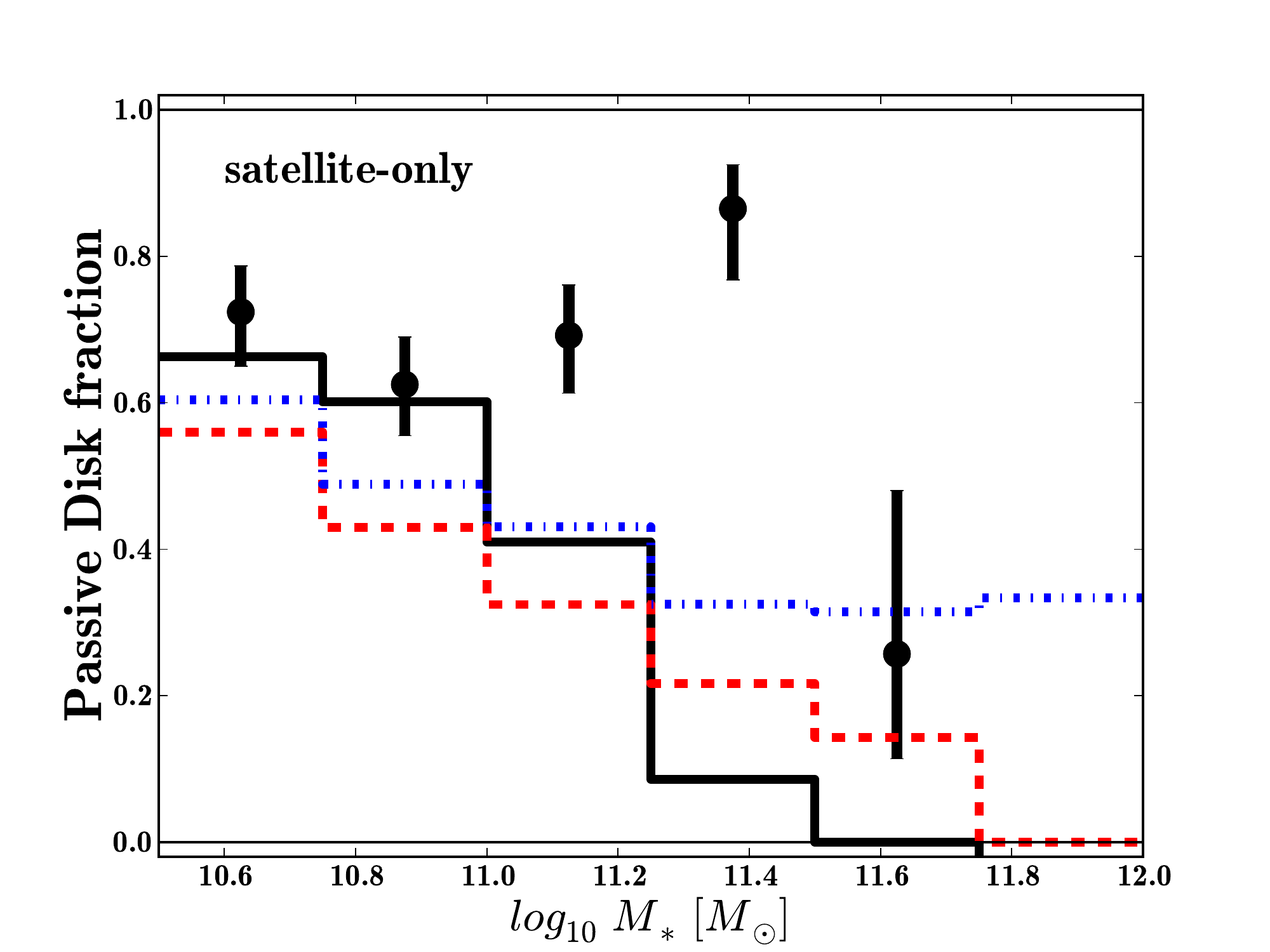}
              \includegraphics[width=0.49\textwidth]{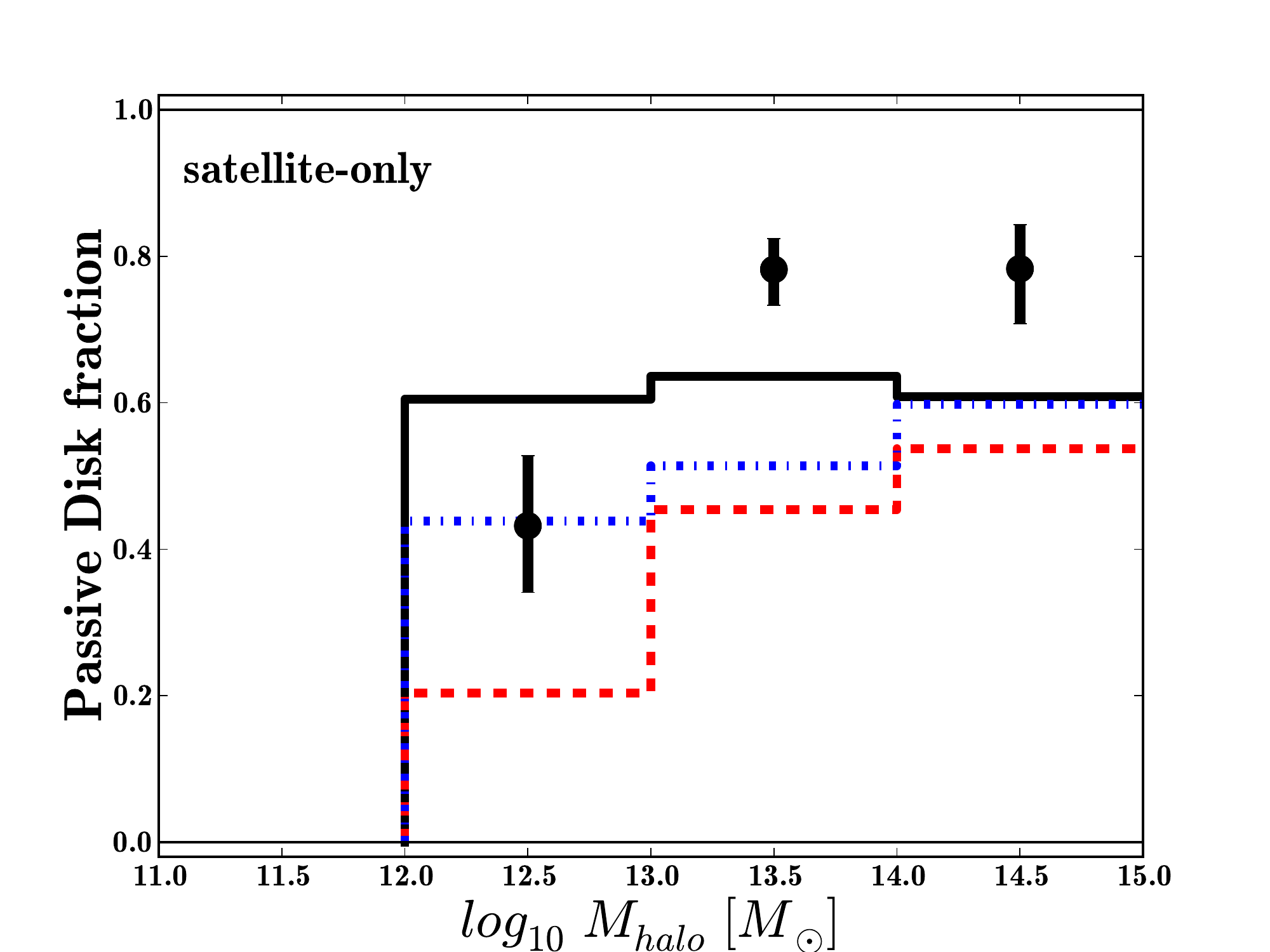}
            }
            \caption{As Figure~\ref{fig:fEz0}, except we now present
              the fraction of passive (SSFR $ < 10^{-11} \yr^{-1}$) disc
              galaxies divided into central and satellite populations
              as a function of stellar and halo mass and contrasting
              observations with the pure mergers model implementations
              (see key).}
  \label{fig:fDpz0}
\end{figure*}

Figure~\ref{fig:fDpz0} shows the same information for passive disc
galaxies. All models produce far too few passive disc galaxies.  This
is especially true for centrals and at high mass. At the centre of
$\Mhalo<10^{12}\Msol$ halos the underproduced passive disc fraction is
due to the overproduction of star-forming disc galaxies, and can be
explained if star formation is in reality more easily suppressed in
such halos, although resolution effects can also be important.  In all
other environments, our models produce roughly the right fraction of
star-forming disc galaxies but too many ellipticals.  Thus, an
underproduction of passive disc galaxies is inevitable. The longer
$\tausat$ version of \morgana\ does not greatly affect the fraction of
passive disc galaxies at the centre of halos.  Instead, it produces
more central star-forming disc galaxies.  Therefore, changing $\tausat$
does not seem to reduce the discrepancy with observations.

\subsection{Total Passive Fraction}\label{sec:passive}

\begin{figure*}
  \centerline{\includegraphics[width=0.49\textwidth]{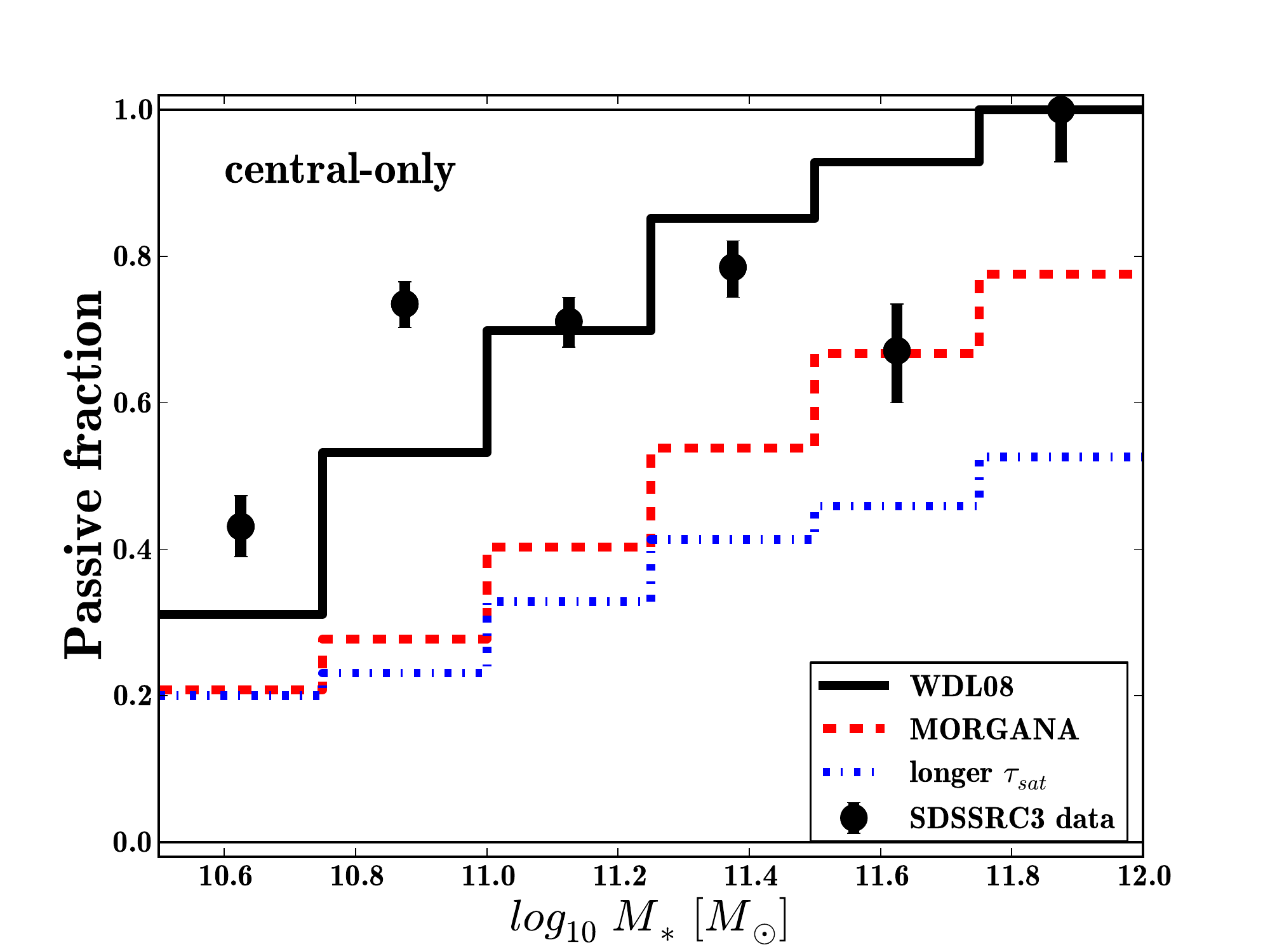}
              \includegraphics[width=0.49\textwidth]{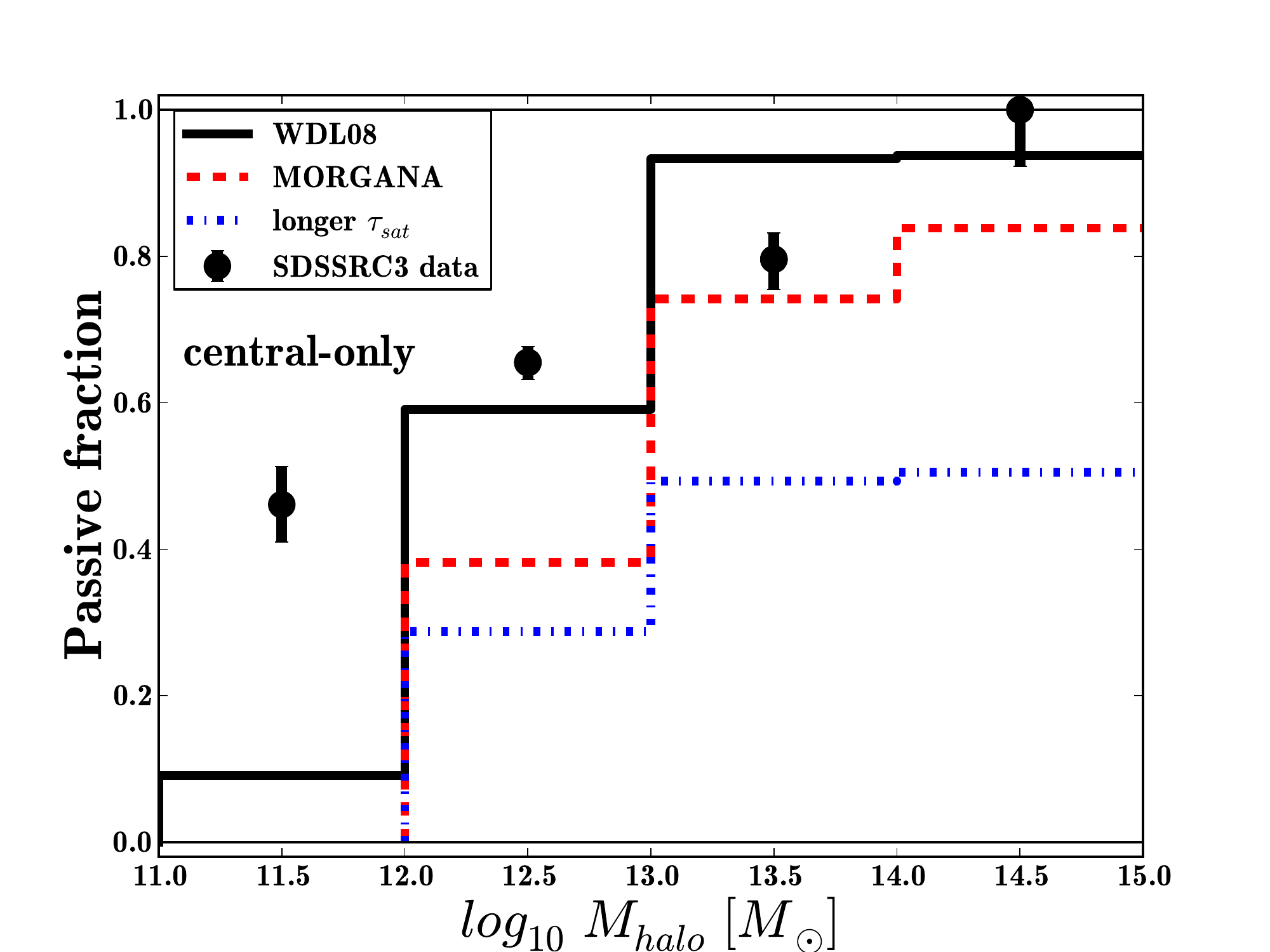}
            }
  \centerline{\includegraphics[width=0.49\textwidth]{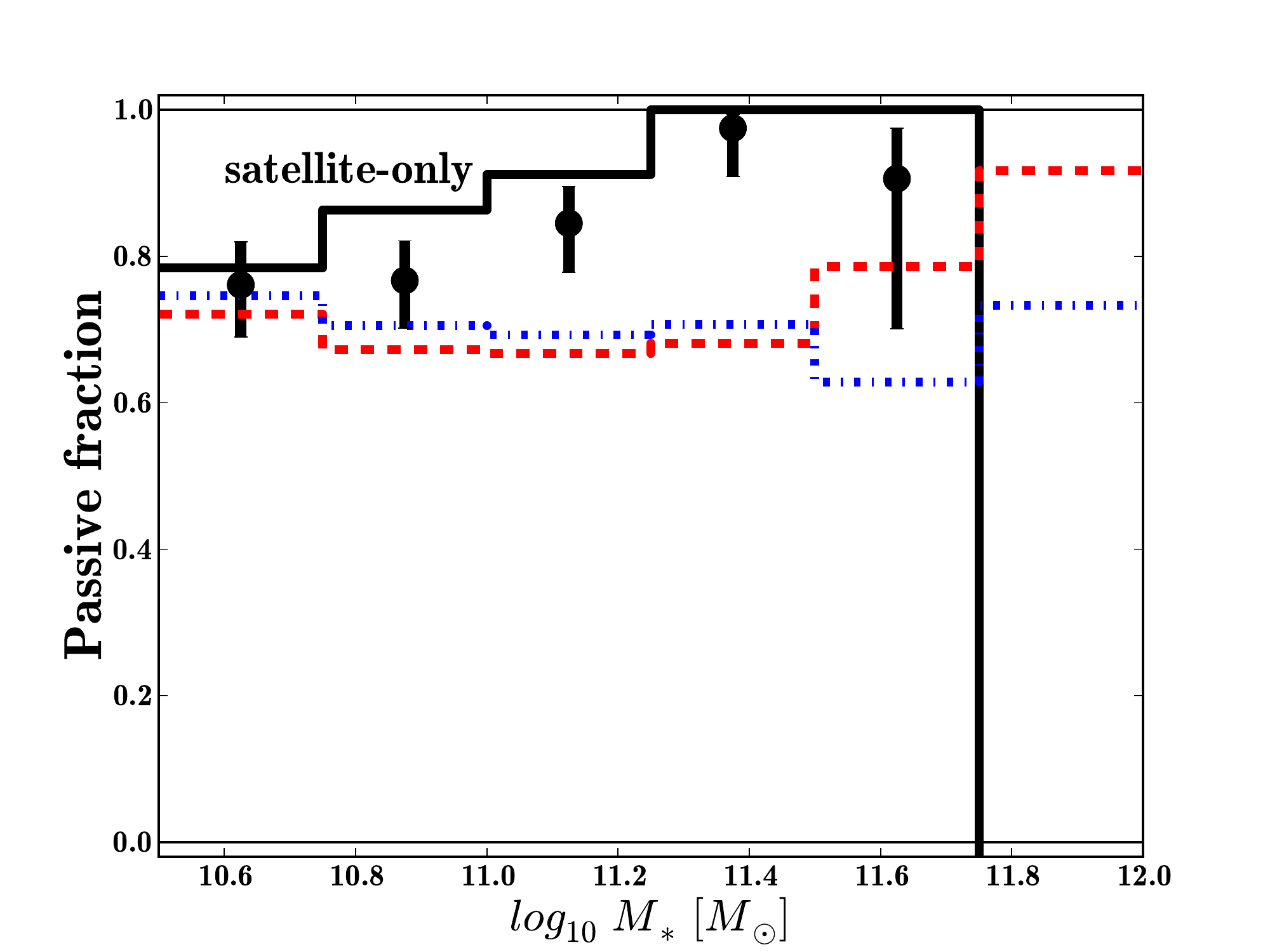}
              \includegraphics[width=0.49\textwidth]{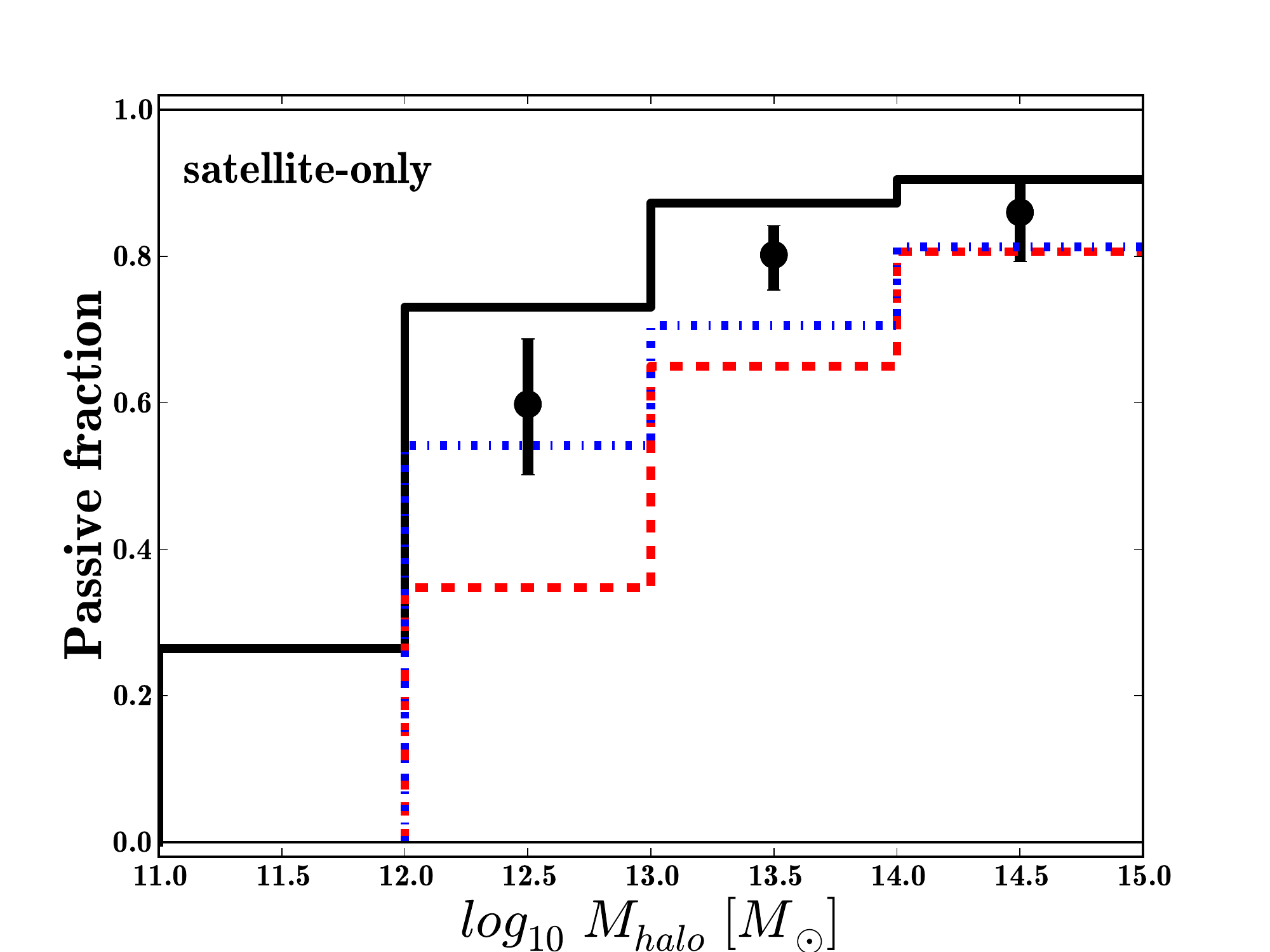}
            }
            \caption{As Figure~\ref{fig:fEz0} except this time we
              present the total fraction of passive (SSFR $\geq
              10^{-11} \yr^{-1}$) galaxies, divided into central and
              satellite populations as a function of stellar and halo
              mass and contrasting observations with the pure mergers
              model implementations (see key).}
  \label{fig:fpz0}
\end{figure*}

We have seen that the fraction of star-forming disc galaxies is
reasonably well reproduced by our models (Figure~\ref{fig:fDsfz0}).  We
have also seen that ellipticals are overproduced by the models at the
expense of passive disc galaxies. Therefore our models produce the
correct total passive (or star-forming) fraction of galaxies, but get
the $\bt$ distribution of passive galaxies wrong: i.e. too many are
converted into ellipticals. To see this more explicitly, we examine the
{\it total passive fraction} of galaxies (with no selection on $\bt$)
in Figure~\ref{fig:fpz0}.

Figure~\ref{fig:fpz0} shows that the total passive fractions of both
central and satellite galaxies are generally in good agreement with the
models. The largest discrepancy is the underproduction of passive
galaxies at the centre of low mass $\Mhalo<10^{12}\Msol$ halos,
previously noted to lead to an overproduction of star-forming disc
galaxies (Section~\ref{sec:pdisc}). Otherwise, our models produce
roughly the right total passive fractions, increasing with stellar and
halo mass. Passive fractions are slightly lower for \morgana\, and
especially with a longer $\tausat$, than for the \citetalias{WDL08}
model: current data appears to slightly favour the higher passive
fractions produced by \citetalias{WDL08}.

Comparing the passive fractions of central galaxies to those of
satellite galaxies indicates that model satellites are more often
passive than centrals of the same mass, particularly at lower stellar
mass. This is due to the modelling of strangulation which assumes
complete and instantaneous stripping of hot gas from satellite galaxies
upon their accretion onto a parent halo. This leads to the quenching of
star formation once the existing cold gas is exhausted.  However, at
the stellar masses we are probing, passive fractions are globally high.
This means our dynamic range to see differences between central and
satellite passive fractions is limited.  Observed fractions have larger
statistical errors, and the satellite passive fraction is only notably
higher than that for centrals in the lowest mass bin
($\Mstellar\leq10^{10.75}\Msol$). Our sample's high stellar mass limit
is likely the main reason why we fail to reproduce the much discussed
overproduction of passive satellite galaxies in group halos by
semi-analytic models \citep[e.g.][]{Font08,Weinmann10} -- this effect
is most clearly seen at lower stellar mass.

\section{Interpretation}\label{sec:interpretation}

Our models create roughly the right number of passive galaxies (in the
stellar mass range probed). However, these passive galaxies too often
have elliptical morphology ($\bt\geq0.7$).

We shall now consider the evidence: When did our model ellipticals
experience their last major merger? What is the ultimate fate of
satellite galaxies? How does quenching of star formation in
  central galaxies proceed, and how is this related to the growth of
  bulges? We shall use these questions to tease out the degrees of
freedom in our models which should ultimately help reconcile the model
population with the observed galaxy population.

\subsection{Hierarchical Growth and the Last Major Merger}\label{sec:hierarchicalgrowth}

Both $\bt$ and star formation rates of galaxies depend sensitively on
their full history of hierarchical growth. The most significant
transformation of a galaxy's morphology happens during a major merger,
and the more recent that merger, the greater the probability that the
galaxy will be observed with elliptical morphology.

\subsubsection{The Last Major Merger}\label{sec:lastmerger}

\begin{figure}
  \includegraphics[width=0.5\textwidth]{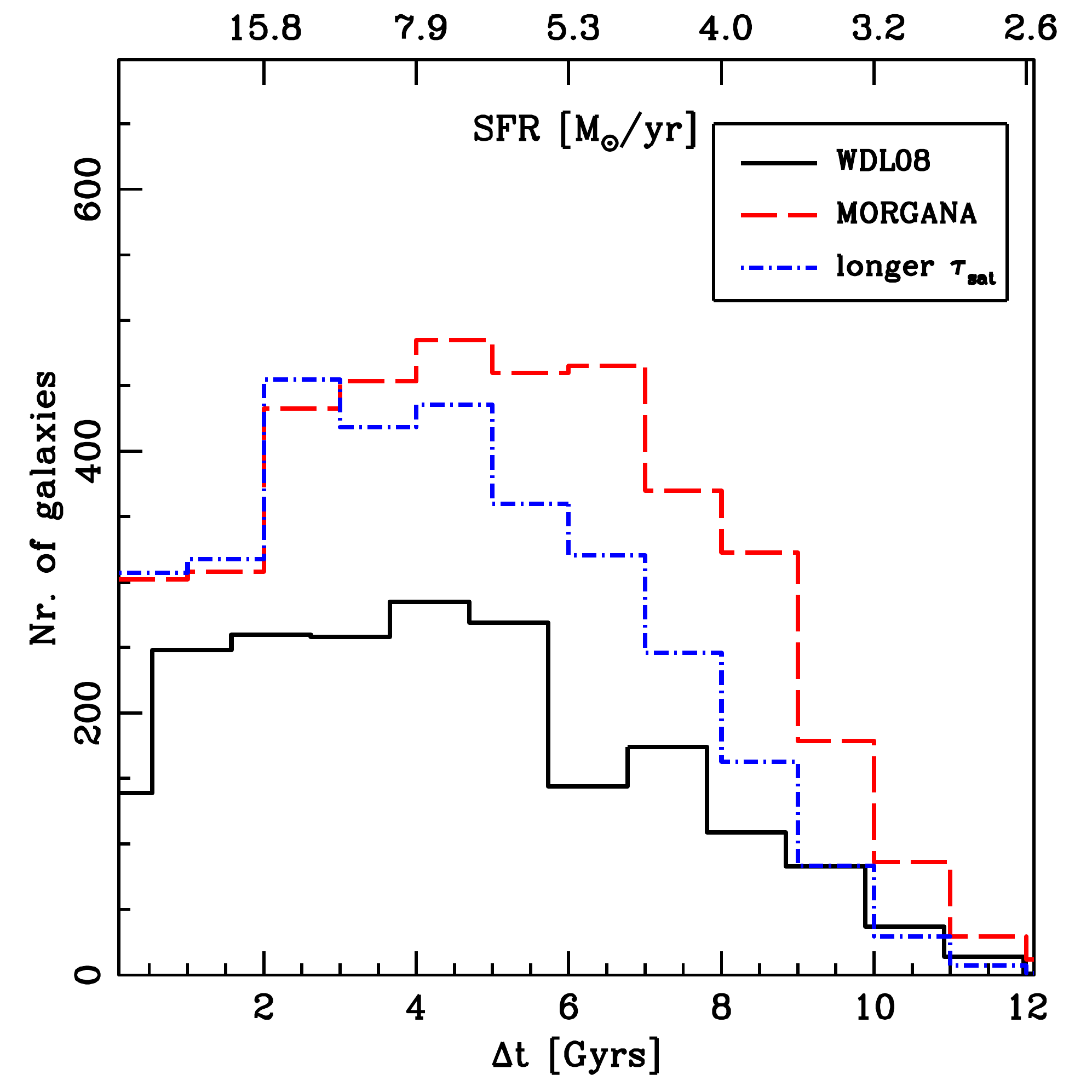}
  \caption{The distribution of the time $\Delta t$ in Gyr since the
    last major merger for all massive ($\Mstellar > 10^{10.5} \Msol$)
    galaxies that are ellipticals at $z=0$ ($\bt\geq0.7$) in our models
    with the pure merger bulge formation implementation.  Black solid /
    red dashed / blue dot-dashed histograms for \citetalias{WDL08},
    \morgana\ and \morgana\ with longer $\tausat$ realizations. The
    upper axis scale shows the average SFR required to form a
    $10^{10.5}\Msol$ disc from the time of the major merger until $z=0$
    -- this is the absolute minimum mass required to form a disc
    dominated $\Mstellar \geq 2\times 10^{10.5} \Msol$ galaxy at
    $z=0$.}
  \label{fig:DTMajorMerger}
\end{figure}

Figure~\ref{fig:DTMajorMerger} shows the distribution of the time
$\Delta t$ in Gyr since the {\it last major merger} for all massive
($\Mstellar > 10^{10.5} \Msol$) galaxies that are ellipticals
($\bt\geq0.7$) at $z=0$ in our models with the pure merger bulge
formation implementations. There are more ellipticals in \morgana\ than
\citetalias{WDL08} which accounts for the different normalization, but
the redshift distribution of the {\it last} major merger for the two
models is very similar. This contrasts with the redshift at which the
ellipticals {\it acquired their morphology} which for most ellipticals
is much higher in \morgana\ than in \citetalias{WDL08} (Figure~9 of
\citetalias{deLucia11}).  Some of the ellipticals formed by \morgana\
which had no recent major merger ($\Delta t \gtrsim 5 \Gyr$) never acquire
an elliptical morphology at all in the version with longer $\tausat$.
This is likely the reason that a longer $\tausat$ leads to more
star-forming disc galaxies: without a major merger, gas continues to
cool and fuel disc growth, while AGN feedback remains inefficient (see
section~\ref{sec:centralquench} for more on AGN feedback and its
relation to merger induced bulge growth).  This is particularly true in
\morgana, where few passive, central disc galaxies are formed
(figure~\ref{fig:fDpz0}).

In all models the bulk of ellipticals experienced their last major
merger at $z<1$ ($\Delta t \lesssim 7.5 \Gyr$).
Table~\ref{table:totalfractions} shows that the global elliptical
fraction is overproduced by our models with pure merger bulge growth
implementations by factors of $2.6$, $4.2$ and $3.7$, respectively, for
\citetalias{WDL08}, \morgana\ and \morgana\ with the longer $\tausat$.
In both models, there have been enough major mergers since $\Delta t
\sim 3.5$--$3.7 \Gyr$ ($z \sim 0.3$) to create all the ellipticals observed
in the $z=0$ Universe! A longer $\tausat$ model does nothing to change
this situation.

It is therefore clear that it is not enough to reduce the major merger
rate at high redshift: {\it The rate of major mergers at low
    redshift is also too high.}

\subsubsection{Post-Merger Disc Regrowth}\label{sec:discregrowth}

Our models maintain the bimodal $\bt$ distribution through relatively
recent elliptical formation plus suppressed cooling via AGN feedback. A
relaxed feedback prescription would allow gas to cool more easily onto
newly formed elliptical galaxies, where it would reform stellar discs. 

On the upper axis of Figure~\ref{fig:DTMajorMerger} we explore the
average disc star formation rates (SFR) which would be necessary for
each elliptical galaxy which has just experienced (its last) major
merger to reform a disc of mass $\Mstellar = 10^{10.5}\Msol$ in $\Delta
t$. This is the limiting case which allows us to transform a $\Mstellar
= 10^{10.5}\Msol$ elliptical galaxy with $\bt=1$ into a $\bt=0.5$
galaxy by $z=0$.

An elliptical formed at $z=0.5$ ($\Delta t \sim 4.9 \Gyr$) requires an
average $\sfr \sim 6.5 \Msolyr$ (initial $\ssfr \sim 0.42 \Gyr^{-1}$). This
is within the scatter of the typical observed $\ssfr \sim 0.3 \Gyr^{-1}$
at that redshift \citep[e.g.][]{Feulner05,Noeske07}. To quote a more
extreme case: to build a $\bt=0.2$ galaxy with $\Mstellar =
10^{11}\Msol$ at $z=0$ we require an average $\sfr = 32 \Msolyr$ since
$z=0.5$, or $\sfr = 16 \Msolyr$ since $z=2$, corresponding to
$\ssfr \sim 0.6 \Gyr^{-1}$ and $\ssfr = 0.32 \Gyr^{-1}$ respectively.  This
is still within reason for a star-forming galaxy, even at $z = 0.5$.

However, we cannot simply relax the feedback prescription and allow
this kind of disk regrowth to reduce the elliptical population.  In
Section~\ref{sec:centralquench} we shall examine the tight
relationship between bulge growth and the quenching of star formation,
grounded in theory and observation, and qualitatively present in our
models. This relationship tells us that ellipticals {\it do not}
continue forming stars at a rate typical of galaxies living on the
star-forming sequence. And indeed, our observations tell us that we
need to form more {\it passive disc galaxies}: the models already form
enough star-forming ones. As an extension to this, any elliptical
galaxy which slowly regrows its disc will spend a long period of time
with $\bt>0.5$. As we have seen in Figure~\ref{fig:fd11comp} (also see
\citet{Weinzirl09,Laurikainen10}), such galaxies are rare -- most star
forming galaxies have $\bt<0.5$.  Finally, even with the $\sfr$ of a
typical star-forming galaxy, an elliptical will not grow a massive
enough disc in the time since $z=0.3$ (since when there are
  enough major mergers to form the entire observed elliptical
  population, section~\ref{sec:hierarchicalgrowth}). Thus, disk
  regrowth cannot compensate for the problem of too many
  elliptical-forming major-merger events, particularly at low
  redshift.

\subsubsection{Disc Survival in Major Mergers}\label{sec:discsurvival}

In our \hop\ implementation of both models, calibrated to
  numerical simulations, major mergers do not entirely destroy discs
\citep[][see also \citealt{Bournaud07}]{Hopkins09}.  We have shown in
Section~\ref{sec:btmass} that with this implementation, residual
  discs in major merger remnants are typically less than $30\%$ by stellar
mass. This is insufficient to explain the observed high fraction of
disc galaxies which have typically much lower $\bt$
\citep[Figure~\ref{fig:fd11comp}, see also][]{Laurikainen10}.  For
this reason we ignored  this implementation in
later analysis.  However, it is worth noting that under the right
circumstances (e.g.  high redshift mergers with high gas fractions) or
with additional physics (e.g.  self-consistent merger-induced heating
of stellar discs; pre-merger stripping -- see
Section~\ref{sec:stellarstripping} -- etc.), a similar implementation
might prove more fruitful.

\subsection{The Fate of Satellites}\label{sec:fatesatellites}

There remains the option of changing the fate of satellite galaxies.
Our models assume that galaxies spend a certain amount of time as
satellites ($\tausat$) and are then accreted onto the central
galaxy; the resulting merger will produce an elliptical if the
  mass ratio is high enough. In a minor merger the bulge growth
  depends upon the mass ratio.   We now consider
various ways in which satellite galaxies might differently evolve
before the merger, leading to a different amount of bulge growth.

\subsubsection{Stellar and Gas Stripping}\label{sec:stellarstripping}

\citet{McCavana12} examine tidal disruption timescales in
dark-matter-only cosmological simulations. These are typically shorter
than timescales for satellite-central coalescence for mass ratios
$\mu\lesssim0.25$. Although stellar profiles are more concentrated
than dark matter, the frequency and mass ratio of minor mergers might
be significantly lower than predicted: using simulations populated
using a sub-halo abundance matching technique \citet{Wetzel10} suggest
that a high fraction of satellite galaxies disrupt into the diffuse
component and may never merge with the central galaxy.

Our ellipticals form predominantly via major mergers. Massive satellites
with short satellite survival times might suffer tidal stripping
(unbinding) of outer disc stars, even while the inner stars survive
\citep{Villalobos12}; this can serve to reduce the mass ratio $\mu$ of
the merger and thus reduce bulge growth. Tidal stripping can operate via
individual, strong interactions with other galaxies, or via multiple
weak interactions with galaxies plus the global halo potential.  The
stripped stars may form an diffuse intrahalo component, or, if stripping
occurs close to the central galaxy, they may eventually be accreted via
a stellar stream.  Alternatively, some fraction of stars may be removed
to the diffuse component {\it during the merger event}, as discussed by
\citet{Monaco06}.\footnote{This is an option in \morgana\ but has not
been applied in our version.} This component would almost certainly be
reaccreted onto the remnant.

Previous work has emphasized that the growth of brightest cluster
galaxies (BCGs) is weaker than predicted by semi-analytic models
\citep{Whiley08,Collins09,Stott10} although recent estimates accounting
for progenitor bias suggest the discrepancy is not as great as
previously thought \citep{Lidman12}. Tidal stripping of satellite
galaxies, which puts stars into the intracluster light component (ICL),
is the proposed solution.

Using stacking analysis \citet{Zibetti05} find $\sim 10\%$ of optical
cluster light is in an intracluster component (ICL), while
\citet{McGee10} find values of $\sim 50\%$ by looking for hostless
supernovae in galaxy groups. In either case, a large fraction of the
stars formed in satellite galaxies can end up in the diffuse medium
and {\it not} in the central galaxy. 

The hot and cold gaseous components of satellite galaxies can also be
stripped via tidal effects and/or ram pressure, suppressing formation
of new stars in the satellite, and reducing mass ratios \citep[see
e.g.][]{Wang08,Zavala12}. As with stellar stripping, the result is less
bulge growth, and thus fewer ellipticals.  Less gas is available to
fuel either a merger-induced starburst, or a new post-merger disc.
While other effects are not included in models, the hot gas is assumed
to be stripped instantaneously when a galaxy is accreted as a
satellite, leading to to a fairly rapid suppression of star formation.
Relative to this, inclusion of other gas stripping effects would not
make much difference to the final mass of the satellite galaxies
\citep[e.g.][]{Lanzoni05}.

{\it To reduce the rate of major mergers at low redshift, tidal
  stripping of stars should be taken into account. } This will reduce
the final population of elliptical galaxies.

\subsubsection{Satellite-Satellite Mergers}\label{sec:satsatmergers}

The \citetalias{WDL08} model tracks the evolution of halos after their
accretion onto a parent halo -- they become subhalos, and their central
and satellite galaxies become subhalo centrals and satellites.
Subhalo-satellite galaxies are able to merge with the subhalo-central
galaxy: in the context of the main halo, these can be regarded as
satellite-satellite mergers; \morgana\ does not consider this process.

\begin{figure}
 \centerline{\includegraphics[width=.5\textwidth]{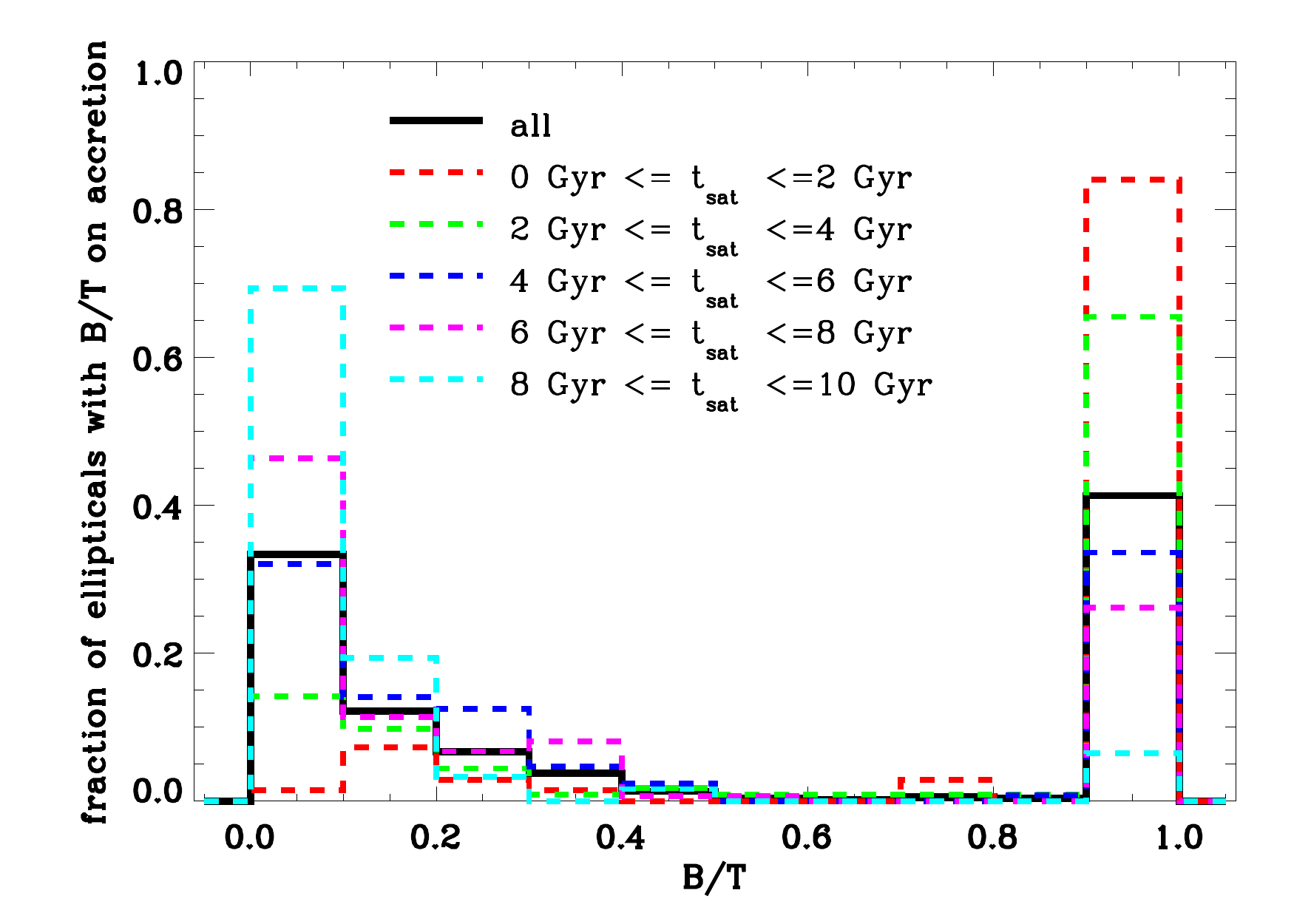}}
  \caption{The distribution of $\bt$ at $\zsat$ -- the time when
    satellite galaxies {\it became} satellites through accretion onto a
    more massive halo -- of all $\bt\geq0.7$ $\Mstellar \geq
    10^{10.5}\Msol$ satellite galaxies at $z=0$ for the WDL08 model
    with the pure mergers implementation (black solid line). This is
    divided into subsets of those which have been satellites for
    different periods, $\tsat$ (coloured dashed lines, see key).  Only
    $\sim 41.4\%$ of \citetalias{WDL08} satellite elliptical galaxies
    ($\bt\geq0.7$) had $\bt\geq0.7$ when they became satellites, with
    $\sim 58.6\%$ having acquired their high $\bt$ since that time --
    this fraction grows with $\tsat$. This must happen in
    satellite-satellite mergers within substructures. }
  \label{fig:Ebtzsat}
\end{figure}

We have taken all satellite elliptical galaxies from the pure merger
implementation of \citetalias{WDL08} with stellar mass $\Mstellar \geq
10^{10.5}\Msol$ at $z=0$. We then ask what was the bulge to total ratio
of their main progenitor at the time that they first became satellites
($\zsat$, i.e. the time at which their halo was accreted onto a more
massive one, and became a subhalo).  Figure~\ref{fig:Ebtzsat} shows the
distribution of $\bt(\zsat)$ (solid, black line).

Only $41.4\%$ of $\bt(z=0)\geq0.7$ satellites had $\bt(\zsat)\geq0.7$.  The
other $58.6\%$ became ellipticals in mergers between satellites. We
examine how this depends upon how long a galaxy has been a satellite,
$\tsat$. These distributions (dashed, coloured lines) clearly show that
those which were more recently accreted are more likely to have
$\bt(\zsat)\geq0.7$, while those which have been satellites for some time
(up to $\sim 10 \Gyr$) are increasingly more likely to have been
accreted as disc-dominated galaxies and to have experienced subsequent
bulge growth. 

The importance of mergers between satellites needs to be better
constrained through realistic cosmological simulations tracing
substructure. Theoretically, one would expect such mergers to take
place in subhalos located in the outer regions of their parent halos
\citep[e.g.][]{Wetzel09}.  However, the evolving tidal effects on the
subhalo and the galaxies within it have not yet been fully explored.

The SDSSRC3 sample is representative, spanning all environments.  The
fraction of satellite galaxies which are ellipticals is low ($<20\%$),
{\it especially} in our most massive halos ($\Mhalo>10^{13}\Msol$).
However, a much larger elliptical fraction ($32\%$) is measured for
$\Mstellar>10^{10}\Msol$ galaxies in a sample of massive clusters from
the WIde-field Nearby Galaxy clusters Survey (WINGS) \citep{Vulcani11}.
The dependence on cluster mass, and significant variance between
clusters \citep[see][]{deLucia12} makes it difficult to constrain the
frequency of mergers between satellite galaxies. Visually identified
major mergers between satellite galaxies living in rich clusters may
help.

Although the direct product of a major satellite-satellite merger is a
{\it satellite elliptical}, this will have a larger mass than its
progenitors and can subsequently merge with the central galaxy with a
shorter dynamical friction timescale and a larger mass ratio.
Nonetheless, the total number of satellite ellipticals is small,
especially in the intermediate mass halos where most major mergers with
the central galaxy take place \citepalias{deLucia11}. Therefore this
will be a secondary effect.

\subsection{Quenching of Star Formation in Central
  Galaxies}\label{sec:centralquench}

As we have seen, part of the problem we face is that we need the
models to produce fewer elliptical galaxies while simultaneously
ensuring that the overall passive fraction (figure~\ref{fig:fpz0}) is
still correctly reproduced. Figure~\ref{fig:fDpz0} shows that the
existing models underpredict passive, central disc galaxies, while
figure~\ref{fig:fDsfz0} shows the abundance of star-forming, central
disc galaxy population is about right.

In practice, we expect that any solution which reduces the central
elliptical abundance will mean a simultaneous increase in the
population of central disc galaxies. For example, tidal stripping
effects can (as suggested in section~\ref{sec:stellarstripping}) reduce
the frequency and mass ratio of mergers, which would diminish the
overall merger history and thus the $\bt$ of remnant central galaxies.
The population of central disc galaxies would then consist of all
existing model central disc galaxies \textit{plus} an extra population
of {\it preserved} central disc galaxies (which in the current models
end up as ellipticals).

If these extra central disc galaxies follow the same pattern as the
existing population of central disc galaxies, then too many of them
will be star-forming and the overall passive fraction will not match
the observations. These preserved central disc galaxies should thus be
predominantly passive. One possible solution flows from the reasonable
hypothesis that these galaxies -- given that they acquire elliptical
morphology in the existing models -- have a richer merger history (and
live in regions which are on average more biased) than is the case for
the existing model central disc galaxies.  Therefore the $\bt$
distribution of these preserved disc galaxies is likely to be biased
towards higher values than is true for the existing disc population.

There is good observational motivation to believe that star formation
is more likely to be quenched in higher $\bt$ galaxies than in low
$\bt$ galaxies.  In the local universe, early-type galaxies with
significant bulges host only low levels of star formation \citep[SFR
$< 1 \Msolyr$,][]{Shapiro10} and cold gas
\citep{Saintonge12,Kauffmann12}.  At higher redshifts, galaxies with
significant bulges (as indicated by higher global S{\'e}rsic indices)
are more likely to be passive \citep{Wuyts11}. Indeed, a significant
bulge may be a requirement for a central galaxy to become passive
\citep{Bell12,Cheung12}.

A correlation between $\bt$ and passive fraction exists in our models
as an indirect consequence of the mergers which drive bulge growth.
Massive galaxies in the centres of massive halos have star formation
quenched when their cold gas reservoir is exhausted by a merger-induced
starburst, and gas cooling at later times is inhibited by
``radio-mode'' AGN feedback. The strength and duty cycle of the
radio-mode heating depend upon the details of the model, and quite
different prescriptions are applied by the \citetalias{WDL08} model
\citep{Croton06} and by \morgana\ \citep{Fontanot06}; a quantitative
comparison of the effects of these different prescriptions on the
properties of host galaxies has been presented in \citet{Fontanot11b}.

\begin{figure}
  \includegraphics[width=0.5\textwidth]{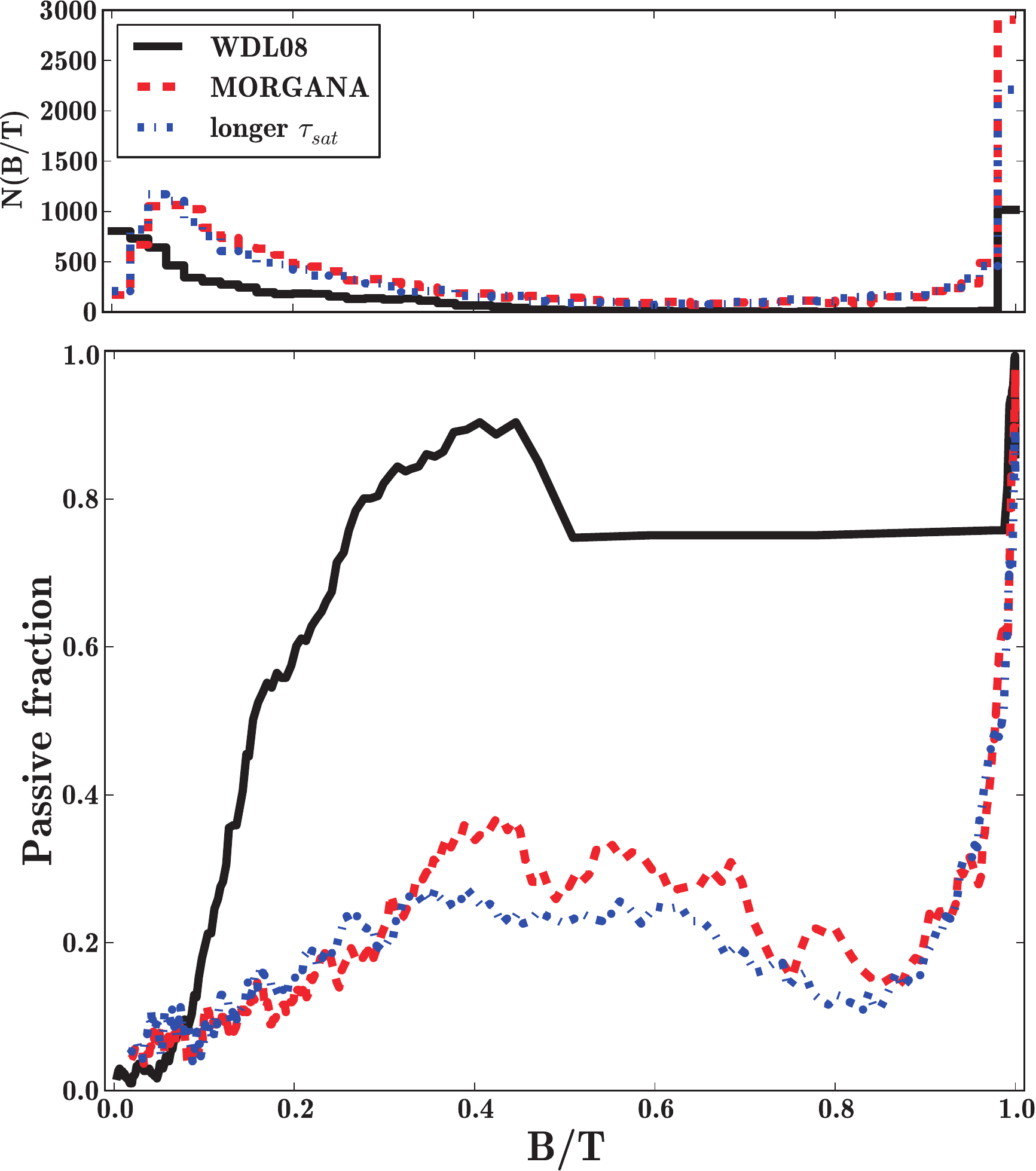}
  \caption{The passive fraction of central galaxies as a function of
    $\bt$ for $\Mstellar\geq10^{10.5}\Msol$ galaxies in the pure
    mergers implementation of \citetalias{WDL08}, \morgana, and the
    longer $\tausat$ version of MORGANA. Fractions are computed in
    running bins of 301 galaxies. For both models, passive fraction is
    a strong function of $\bt$ (see text). The top panel shows the
    distribution of galaxies in $\bt$ for each model.}
  \label{fig:ssfrbt}
\end{figure}

To see how the $\bt$ of a galaxy is correlated in practice with the
quenching of star formation in our models, we examine the relationship
between the fraction of passive central galaxies and their $\bt$ ratio
in figure~\ref{fig:ssfrbt}. The upper panel of this figure displays the
overall distribution of $\bt$ for our model galaxies. The vast majority
of model galaxies are found in two main regimes of $\bt$: one below
$\bt\sim0.4$ and one above $\bt\sim0.95$. Given that we are dealing
with just the pure-mergers versions of our models, the reason for this
distribution is not hard to identify. \textit{Major} mergers
automatically result in remnants with $\bt=1$.  A single minor merger
involving (intially) bulgeless galaxies, on the other hand, cannot
produce a remnant with $\bt$ more than $\frac{0.3}{1.3}\sim0.23$ in
\citetalias{WDL08}, because the bulge mass in the remnant comes
entirely from the satellite galaxy (which, for minor mergers, has mass
ratio $\mu<0.3$). In \morgana\ the resulting bulge mass is supplemented
by rapid star formation in cold gas from the satellite, but this
typically adds little to the final bulge mass at low redshift. So
galaxies with $\bt \ga 0.23$ require a history of multiple significant
minor mergers -- or else a major merger followed by substantial disc
regrowth. But disc regrowth is not significant in our models, as can be
seen by the lack of galaxies with $0.45\lesssim\bt\lesssim0.95$ and the
high passive fraction at high $\bt$.

The main (lower) panel of figure~\ref{fig:ssfrbt} shows how the
fraction of model central galaxies which are passive depends upon
$\bt$. Within each of the two populated regimes, which we shall call
the {\it minor merger regime} ($\bt\lesssim0.4$) and the {\it major
  merger regime} ($\bt\gtrsim0.95$), the fraction of central galaxies
which are passive increases with $\bt$ in both models. However, for
$\bt \ga 0.1$, disc galaxies are quenched much more efficiently in the
\citetalias{WDL08} model than in \morgana.  This is consistent with
figure~\ref{fig:fDpz0}, which shows that \citetalias{WDL08} produces
more passive disc galaxies than \morgana\ in the halo mass range
$\Mhalo\sim10^{12-14}\Msol$.

The challenge will be to adapt the merger and/or star formation
histories of our galaxies such that many no longer become ellipticals,
while still retaining the suppression of star formation which goes
along with those mergers. The likelihood that preserved central disc
galaxies will tend to have moderate-to-high $\bt$ values -- due, as
suggested above, to their residing predominantly in halos with rich
merger histories -- may help in this regard. As figure~\ref{fig:ssfrbt}
indicates, however, this is a more plausible solution for the
\citetalias{WDL08} model: if the adapted central disc galaxies are
assumed to have (for example) $\bt = 0.3$, then $\sim80\%$ of these
will be passive in the \citetalias{WDL08} model, but only $\sim20\%$
will be passive in \morgana. The \morgana\ model clearly requires more
efficient suppression of star formation in the minor merger regime for
central disc galaxies.

\section{Summary and Prospects}\label{sec:conclusions}

We have presented the morphological composition of central and
satellite galaxy populations at $z=0$ for both observed and
semi-analytic mock samples.  We divide our samples into elliptical and
disc galaxies, and also into passive and actively star-forming
galaxies, which means we are effectively examining the co-evolution of
bulge to total ratio ($\bt$) and star formation rate.  To understand
the role of galaxy mergers, we concentrate primarily on the {\it pure
  mergers} bulge formation implementations, applied to both
\citetalias{WDL08} and \morgana\ models, as presented by
\citetalias{deLucia11}.

Analysis of $\bt$ as a function of $\Mstellar$ shows that the
pure-mergers bulge-formation implementation produces two peaks: one at
$\bt=1$ and one at $\bt=0$ with a significant tail up to $\bt\sim0.6$;
this is consistent with recent decompositions of local galaxies
\citep{FD11}.  Our alternative \hop\ implementations -- based on the
simulations of \citet{Hopkins09} and described by
\citetalias{deLucia11} -- produce almost no galaxies with $\bt=1$;
instead, they produce a significant number of galaxies with $0.7
\lesssim B/T < 1.0$, more than are seen in the nearby universe.

For a reference morphological catalogue of nearby galaxies, we used the
SDSSRC3 sample described by \citetalias{Wilman12}, setting a stellar
mass limit at $\Mstellar=10^{10.5}\Msol$. Significantly, this catalogue
separates ellipticals from S0 galaxies, enabling us to identify
galaxies hosting significant discs.  We compared the observed fraction
of elliptical galaxies with the fraction of $\bt\geq0.7$ model
galaxies, and the observed fractions of disc (S0 or spiral) galaxies,
subdivided into passive and star-forming, with the equivalent model
disc galaxies ($\bt<0.7$). We have defined passive galaxies to be those
with specific star formation rates below $\ssfr = 10^{-11} \yr^{-1}$; the
rest we consider to be star-forming. To examine the imprint of
hierarchical growth, we have studied the fraction of each morphological
type separately for central and satellite galaxies, and as a function
of both stellar and halo mass.

Both models get the total fraction of passive galaxies, and the
fraction of star-forming disc galaxies, about right for $\Mstellar\geq
10^{10.5}\Msol$ galaxies as a function of stellar and halo mass, for
both central and satellite galaxies. The only exception is for halos
of mass $\Mhalo<10^{12}\Msol$, where both models overproduce star
forming disc galaxies at the centre of halos. In our models, cooling
and accretion of star-forming gas is very efficient in halos of this
mass, and heating sources are too weak to quench the cooling
flow. Resolution effects also impact our ability to trace the merger
trees of such halos.

The model elliptical fraction increases with stellar mass for both
central and satellite galaxies, but a strong increase with halo mass is
only seen for central galaxies. This is in {\it qualitative agreement}
with observations, consistent with the picture that ellipticals are
predominantly formed at the centre of halos, and that their formation
tracks the hierarchical growth of the halo, and the stellar mass of the
galaxy. 

Despite this success, both models {\it overproduce elliptical galaxies}
by a factor of a few. They do this at the expense of passive disc
galaxies. I.e., {\it while the models get the passive fraction about
  right in the stellar mass range studied, far too many of these become
  ellipticals with $\bt\sim1$}.

This is not highly sensitive to any potential misclassification of
satellite galaxies as centrals, because there is little difference
between central and satellite morphological fractions {\it at fixed
  stellar mass}.

Based on our work, we can identify two requirements for the evolution
of central galaxies which should be met by an improved model.  First,
a majority of the galaxies which (in the models) currently undergo
major mergers and become ellipticals would need to either retain or
reform a significant ($\gtrsim50\%$) disc component. Second, star
formation in these galaxies must nonetheless be quenched.

To reduce the formation of ellipticals, we have considered several
options.  The \hop\ bulge growth implementations yield residual discs
which survive major mergers, but these are too low mass relative to the
bulge (so that $\bt\gtrsim0.7$). Post-merger regrowth of discs would
produce too many present-day star-forming disc galaxies with
significant bulges ($\bt \gtrsim 0.5$), rather than the required
passive disc population. Increasing the survival time for satellites in
\morgana\ does reduce the elliptical population, but this also
increases the population of star-forming disc galaxies instead of the
passive disc fraction.

Most ellipticals in both models experienced their last major merger
after z $\sim 1$. Simulations suggest it is feasible for both gas and
stars to be stripped from many satellite galaxies before they merge
with the parent halo’s central galaxy
\citep[e.g.][]{McCavana12,Villalobos12}. This would lead to a
reduction of the merger mass ratio, and thus yield more minor mergers
instead of major mergers for central galaxies, especially at lower
redshifts where more galaxies live in more massive halos.

Current semi-analytic models (including ours) overproduce low mass
galaxies at $z\gtrsim0.5$
\citep{Fontana06,Fontanot07,LoFaro09,Marchesini09,Fontanot09,Weinmann12}.
While we have as yet no working solution to this problem, it may effect
our predictions: i.e. the star formation history of low mass galaxies
has implications for the mass function of satellites as a function of
time, and therefore on the rate and mass ratio of mergers.

Examination of the correlation between $\bt$ and passive fraction shows
that both models are increasingly efficient at quenching star formation
in central galaxies as $\bt$ increases (in the $\bt < 0.4$ regime where
most disc galaxies are found). If changes which produce more disc
galaxies instead of ellipticals also tend to produce disc galaxies with
significant bulges (e.g., $\bt \sim 0.3$) -- a plausible supposition,
given that these galaxies currently have significant merger histories,
and turning major mergers into minor mergers will still yield disc
galaxies with significant bulges -- then many of these should still be
passive, something which is needed to match the observations. Since
\morgana{} is significantly less effective at quenching disc galaxies
than \citetalias{WDL08}, the \morgana{} models will still require more
efficient quenching of star formation in the $\bt \la 0.5$ regime than
they currently achieve.

We conclude that as models of bulge growth and the quenching of star
formation are inextricably intertwined, a physical model must get both
$\bt$ and star formation rates correct.  In reality, $\bt$ is
difficult to define observationally, and so progress can be made by
considering galaxies at the extreme ends of the $\bt$ distribution:
bulgeless galaxies \citep{Fontanot11} and elliptical galaxies (this
paper). A complete picture of the quenching of star formation in
central galaxies requires matching the observed dependence of the
passive galaxy fraction on $\Mstellar$, $\Mhalo$ and $\bt$.

\section*{Acknowledgments}

DW acknowledges the support of the Max Planck Gesellschaft.  FF
acknowledges financial support from the Klaus Tschira Foundation and
the Deutsche Forschungsgemeinschaft through Transregio 33, ``The Dark
Universe''. Some of the calculations were carried out on the ``Magny''
cluster of the Heidelberger Institute f\"ur Theoretische Studien.  GDL
acknowledges financial support from the European Research Council under
the European Community's Seventh Framework Programme
(FP7/2007-2013)/ERC grant agreement n.\ 202781.  PE was supported by
the Deutsche Forschungsgemeinschaft through Priority Programme 1177
``Galaxy Evolution''. PM has been partially supported by a Fondo
di Ricerca di Ateneo grant of Universit\`a di Trieste.

Funding for the creation and distribution of the SDSS Archive has been
provided by the Alfred P. Sloan Foundation, the Participating
Institutions, the National Aeronautics and Space Administration, the
National Science Foundation, the U.S. Department of Energy, the
Japanese Monbukagakusho, and the Max Planck Society.  The SDSS Web site
is http://www.sdss.org/.

The SDSS is managed by the Astrophysical Research Consortium (ARC) for
the Participating Institutions.  The Participating Institutions are The
University of Chicago, Fermilab, the Institute for Advanced Study, the
Japan Participation Group, The Johns Hopkins University, the Korean
Scientist Group, Los Alamos National Laboratory, the
Max-Planck-Institute for Astronomy (MPIA), the Max-Planck-Institute for
Astrophysics (MPA), New Mexico State University, University of
Pittsburgh, University of Portsmouth, Princeton University, the United
States Naval Observatory, and the University of Washington.

We would like to thank Alessandra Beifiori for comments which helped to
improve the manuscript and the many others with whom we have held
interesting discussions about the results. We thank the anonymous
referee for their interest and comments.

\bibliography{ms}

\appendix

\section{The standard models}\label{sec:stdmodels}

In the main body of this paper, we have described how the fraction of
elliptical, passive disc and star-forming disc galaxies depend on both
stellar and halo mass, for central and for satellite galaxies.  The
main result is that the fraction of visually classified elliptical
galaxies in the real Universe is significantly lower than the fraction
of ellipticals defined to have $\bt\geq0.7$ in two independent
semi-analytic models of galaxy formation, \citetalias{WDL08} and
\morgana, and with a implementation which forms bulges only during
mergers of galaxies (the {\it pure mergers} model). Instead, there is a
higher fraction of real galaxies which are passive
(SSFR $< 10^{-11} \yr^{-1}$) but still possess discs (including both S0s
and spirals).

\begin{figure}
  \centerline{\includegraphics[width=0.24\textwidth]{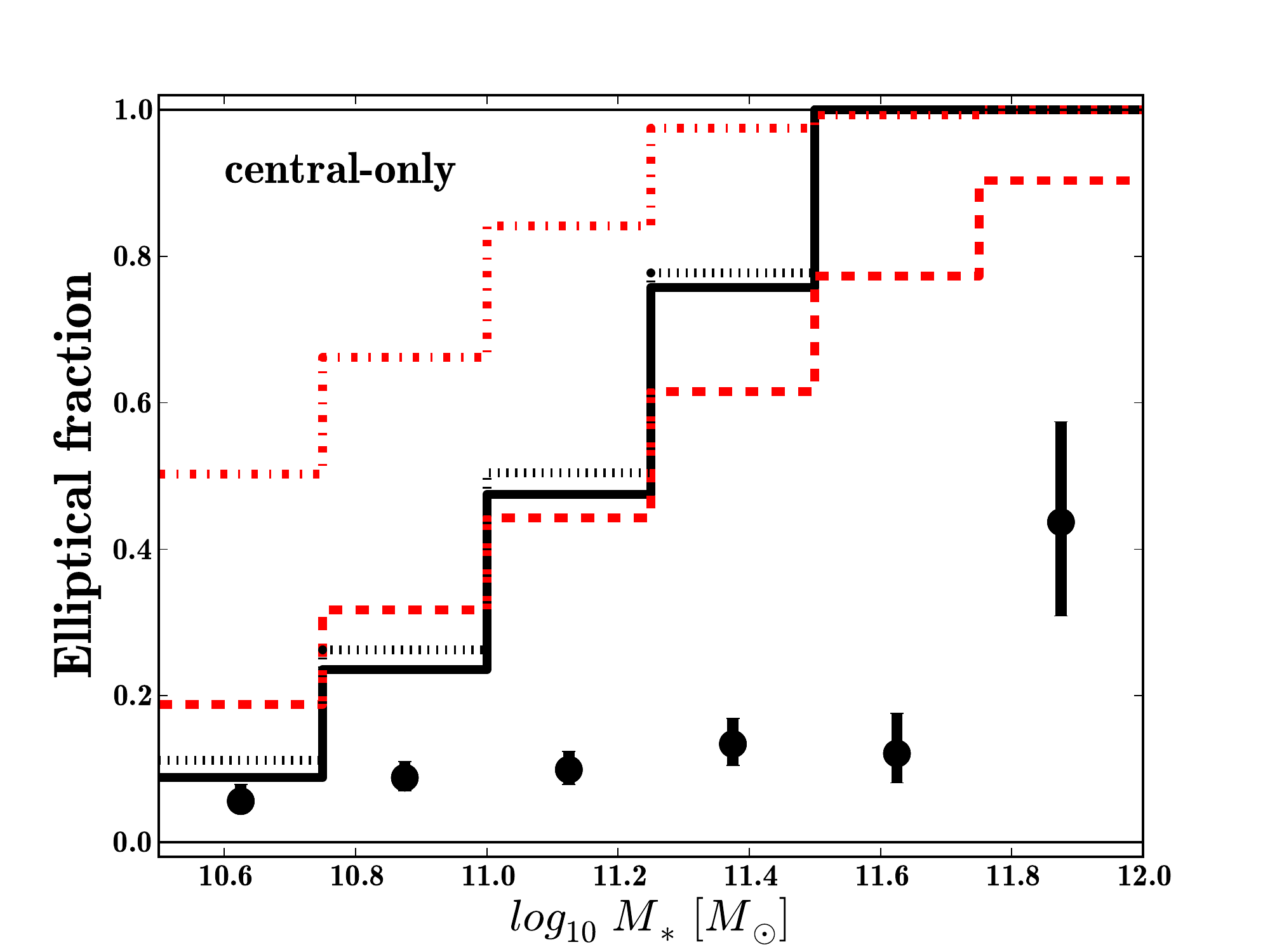}
              \includegraphics[width=0.24\textwidth]{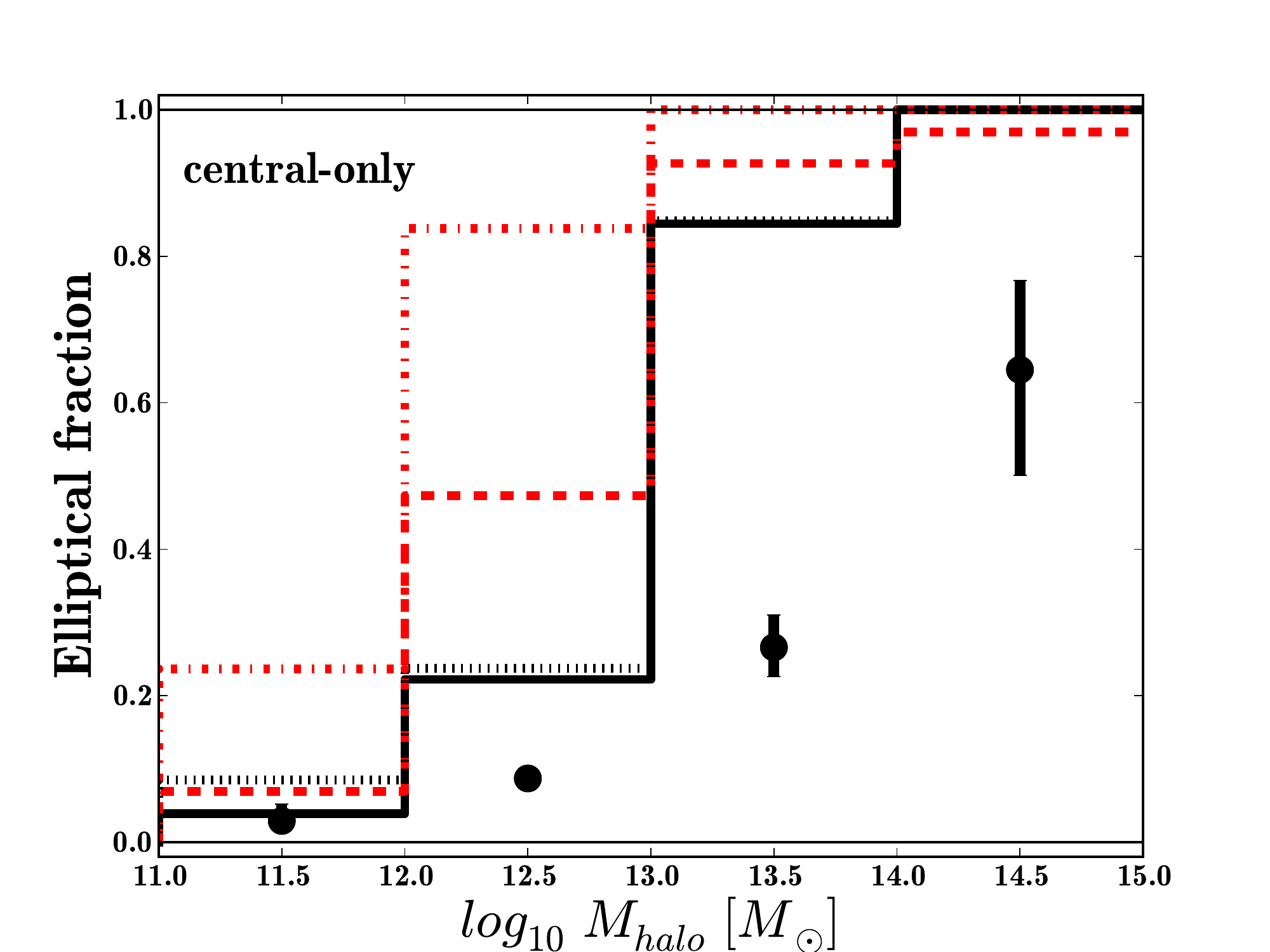}
            }
  \centerline{\includegraphics[width=0.24\textwidth]{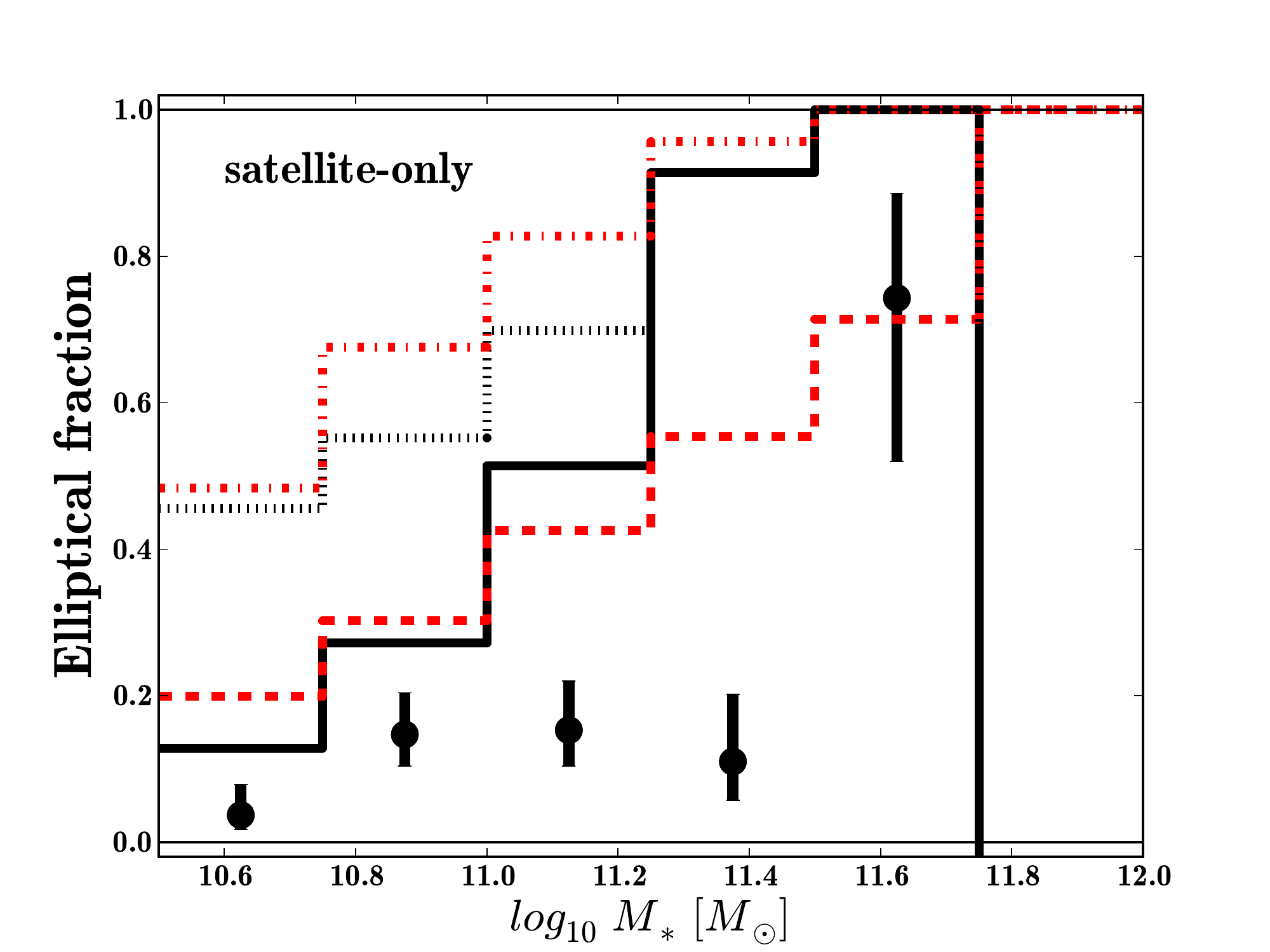}
              \includegraphics[width=0.24\textwidth]{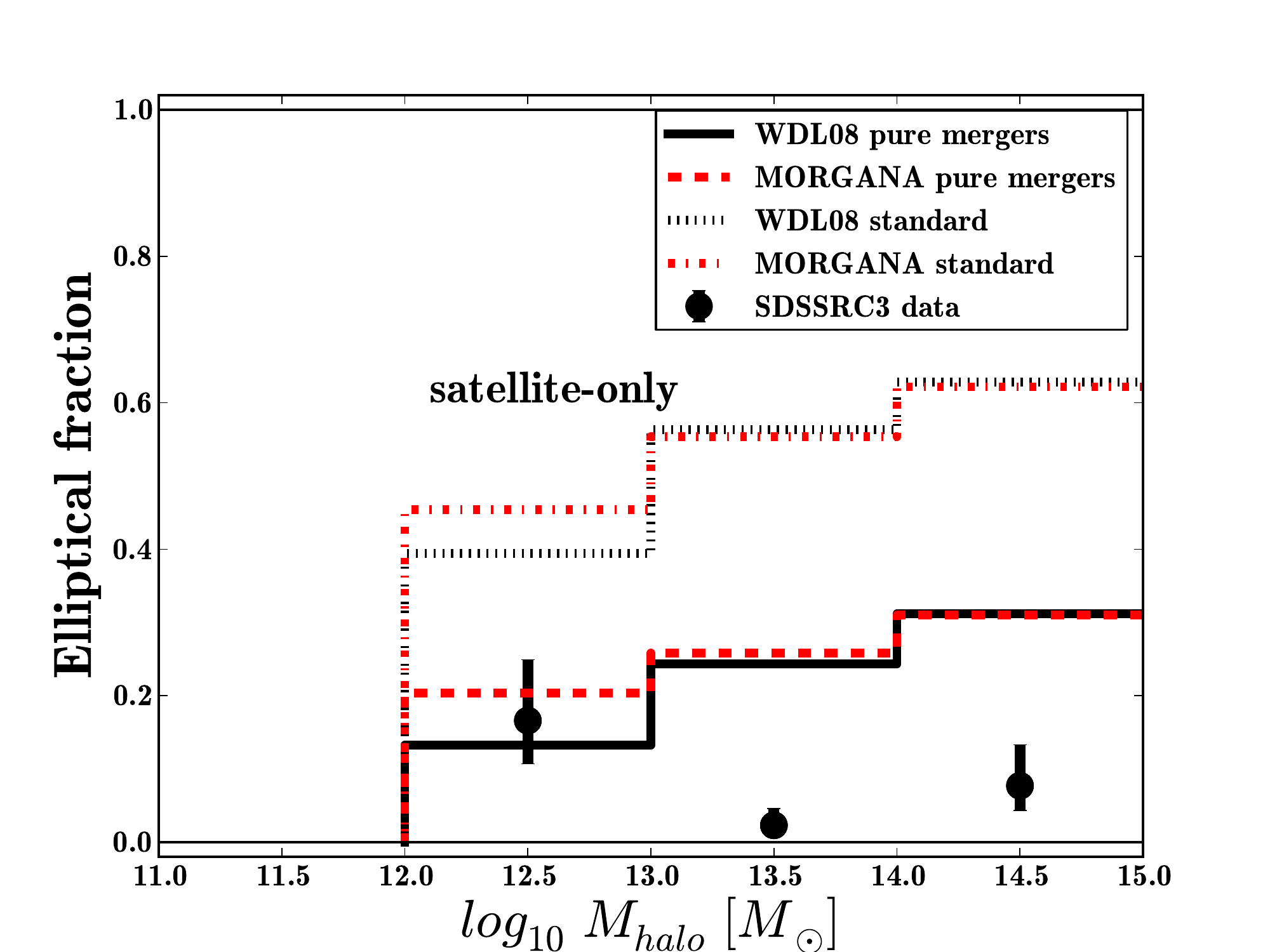}
            }
            \caption{Elliptical galaxy fraction (black points, SDSSRC3
              sample) for $\Mstellar \geq 10^{10.5}\Msol$ galaxies. In
              the top row we only consider central galaxies and in the
              bottom row we only consider satellite galaxies. This is
              compared with the fraction of model elliptical galaxies
              ($\bt\geq0.7$) in the pure mergers implementations of
              \citetalias{WDL08} and \morgana\ models (solid black and
              dashed red lines respectively, as in
              Figure~\ref{fig:fEz0}) and the standard versions of those
              models (including disc instabilities, dot-dashed black
              and dotted red lines).}
  \label{fig:fEz0std}
\end{figure}

\begin{figure}
  \centerline{\includegraphics[width=0.24\textwidth]{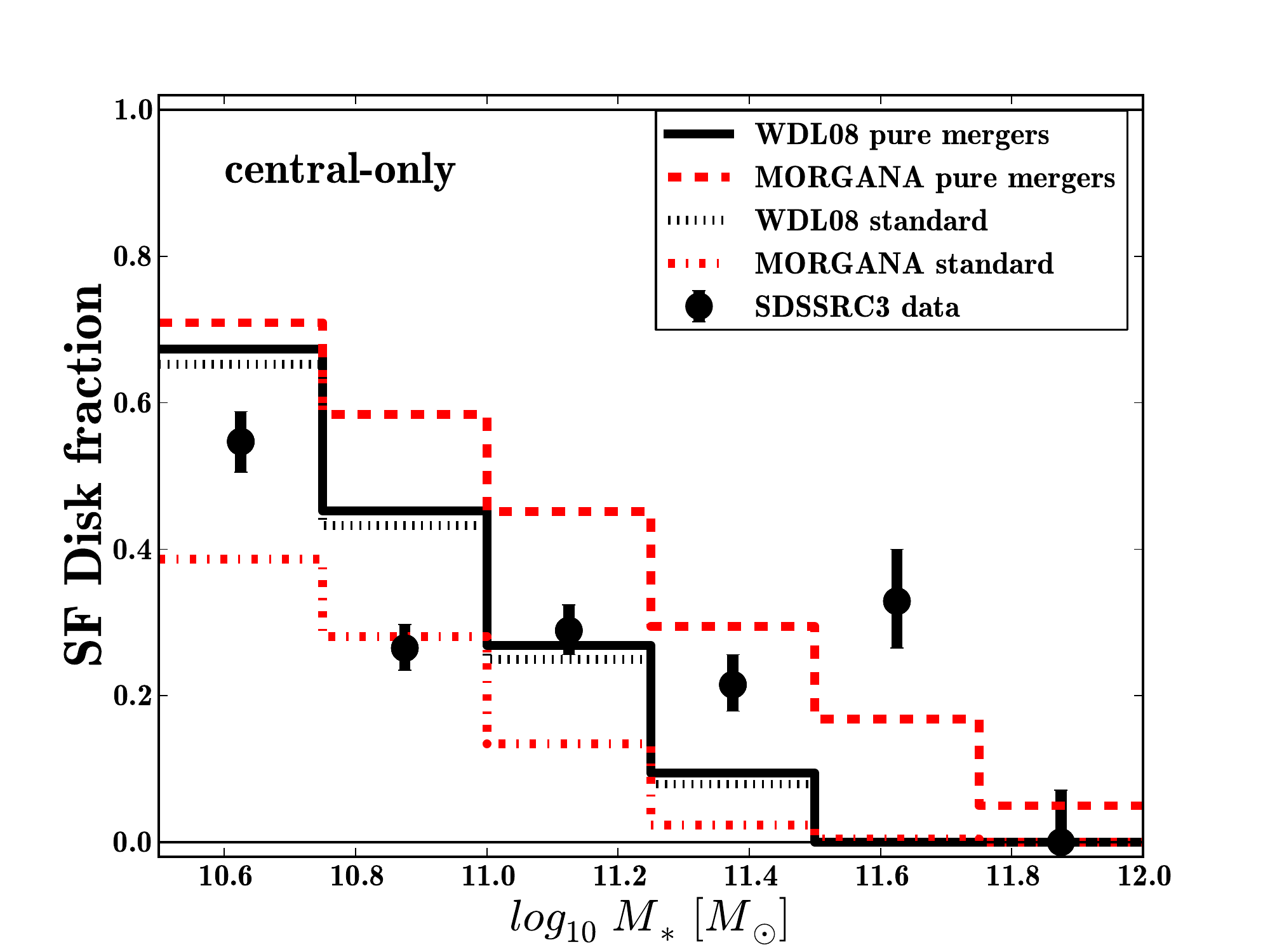}
              \includegraphics[width=0.24\textwidth]{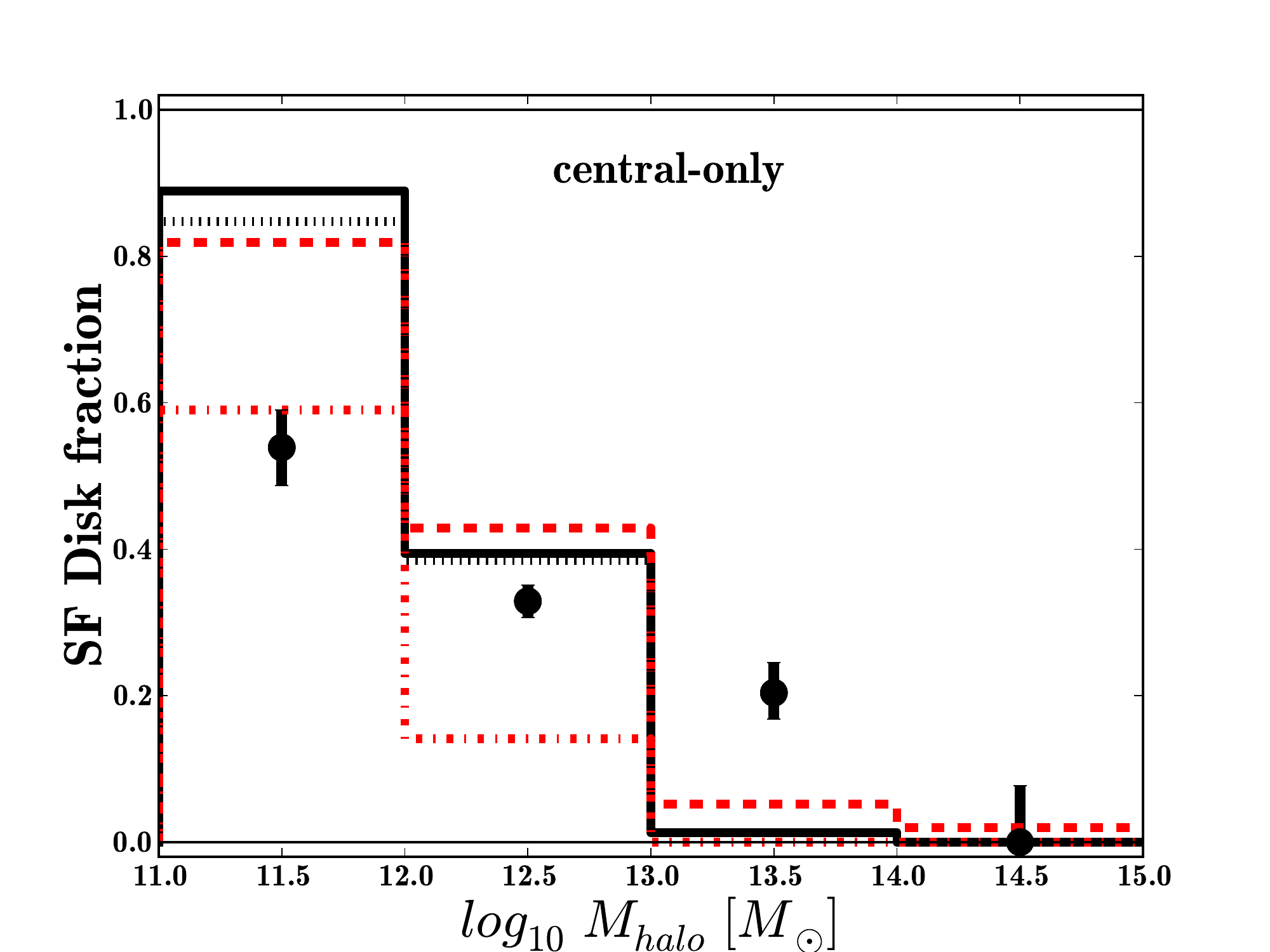}
            }
  \centerline{\includegraphics[width=0.24\textwidth]{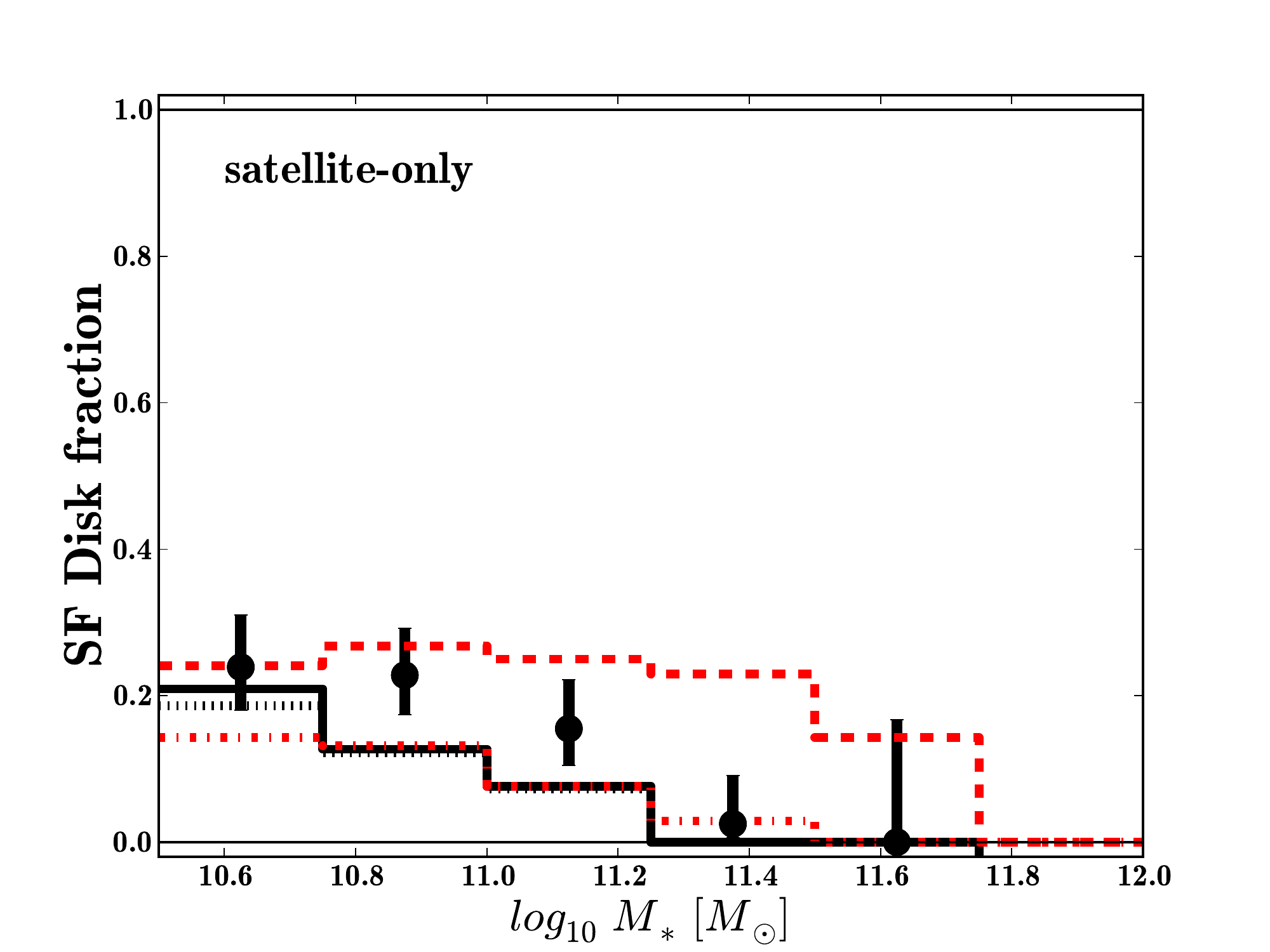}
              \includegraphics[width=0.24\textwidth]{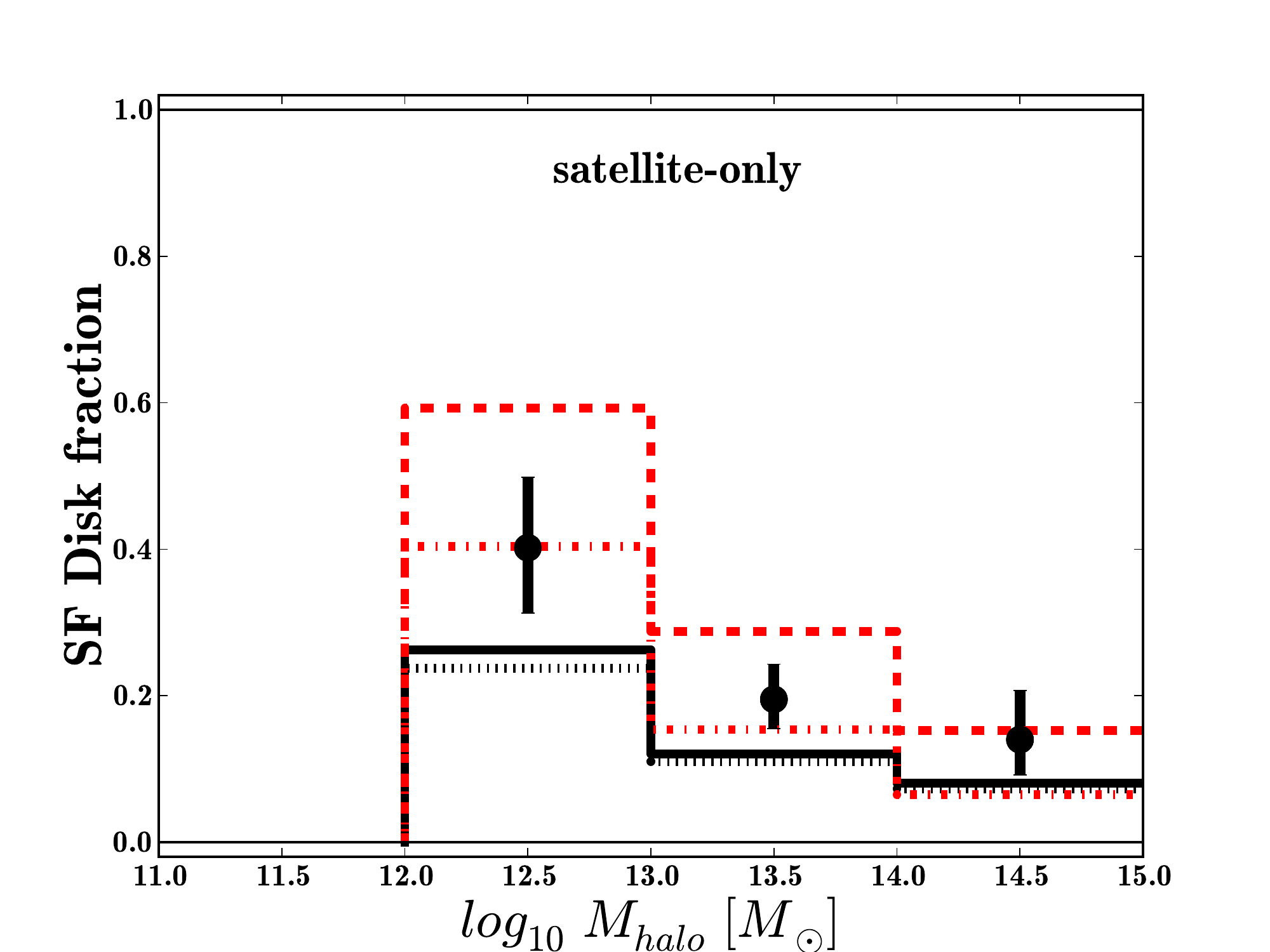}
            }
            \caption{star-forming disc galaxy fraction. Otherwise as
              Figure~\ref{fig:fEz0std}. }
  \label{fig:fDsfz0std}
\end{figure}

\begin{figure}
  \centerline{\includegraphics[width=0.24\textwidth]{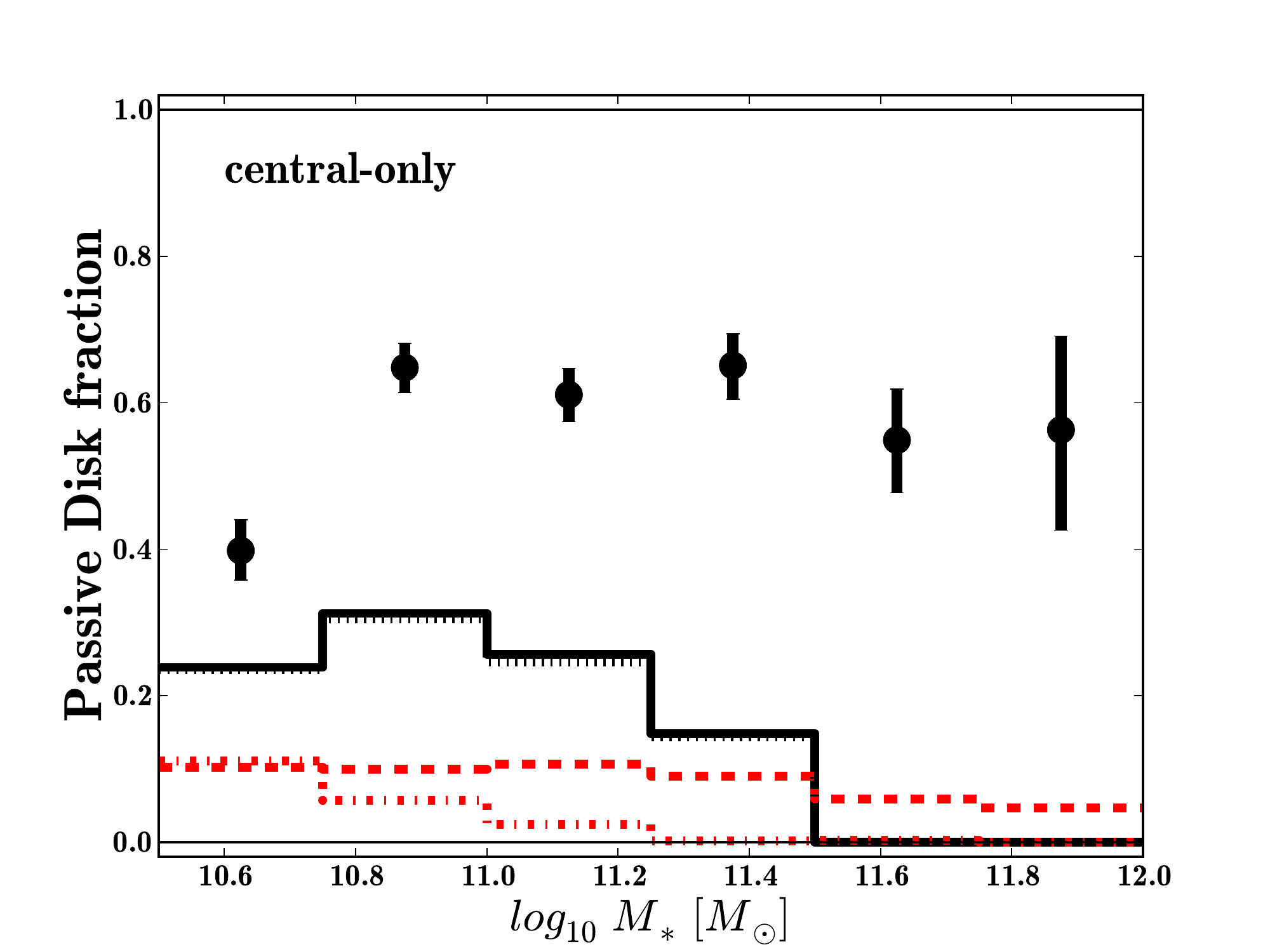}
              \includegraphics[width=0.24\textwidth]{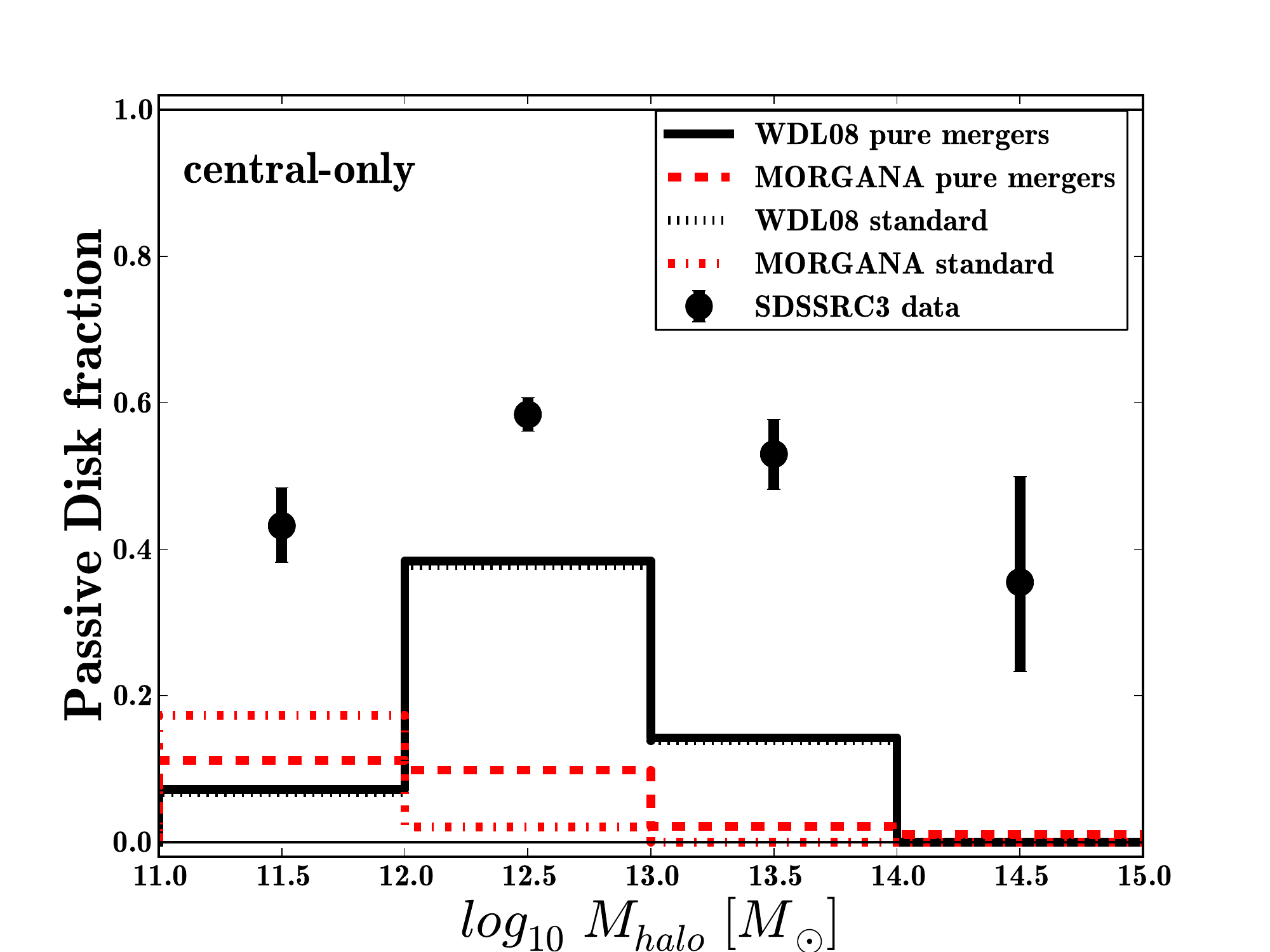}
            }
  \centerline{\includegraphics[width=0.24\textwidth]{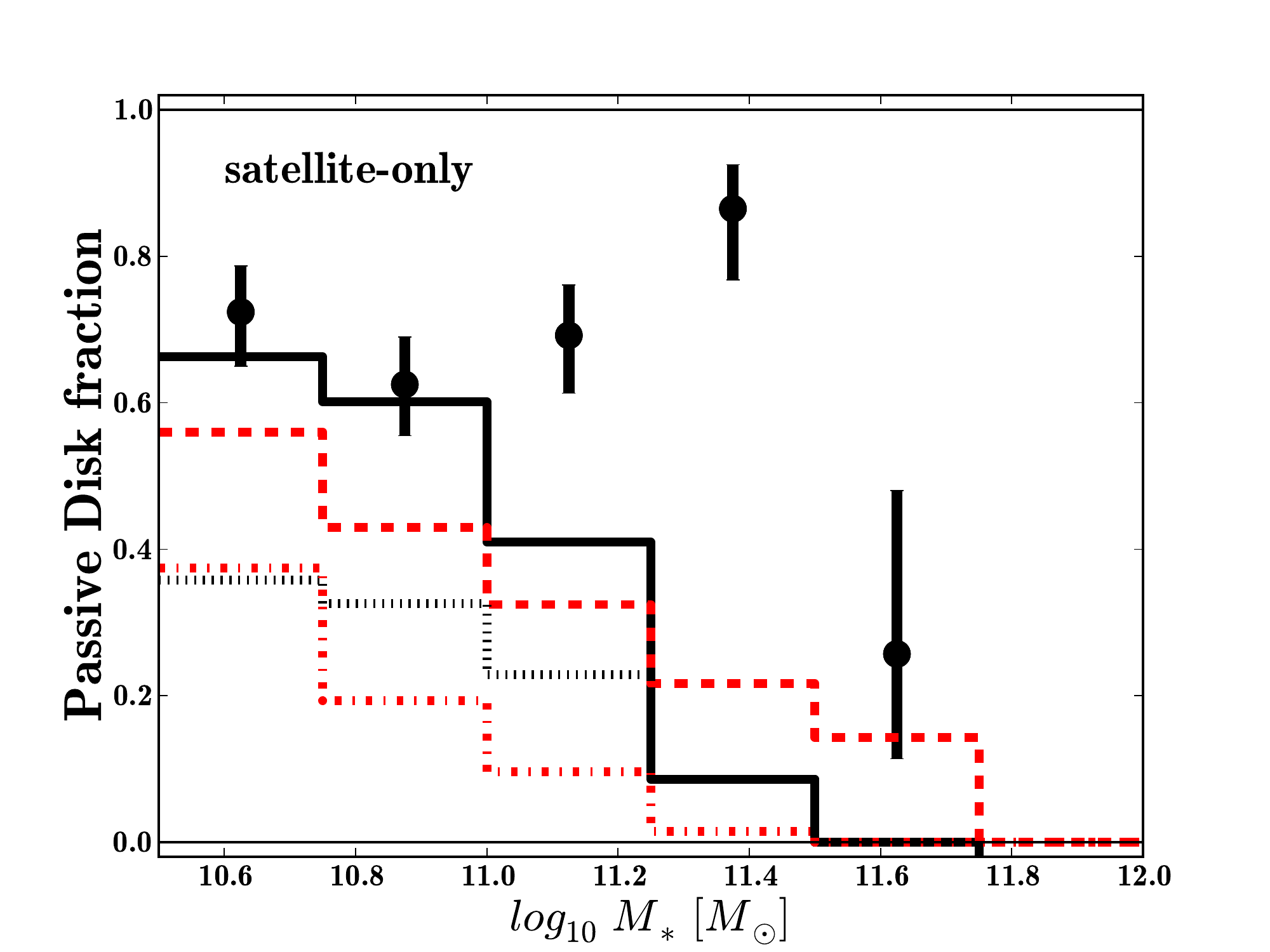}
              \includegraphics[width=0.24\textwidth]{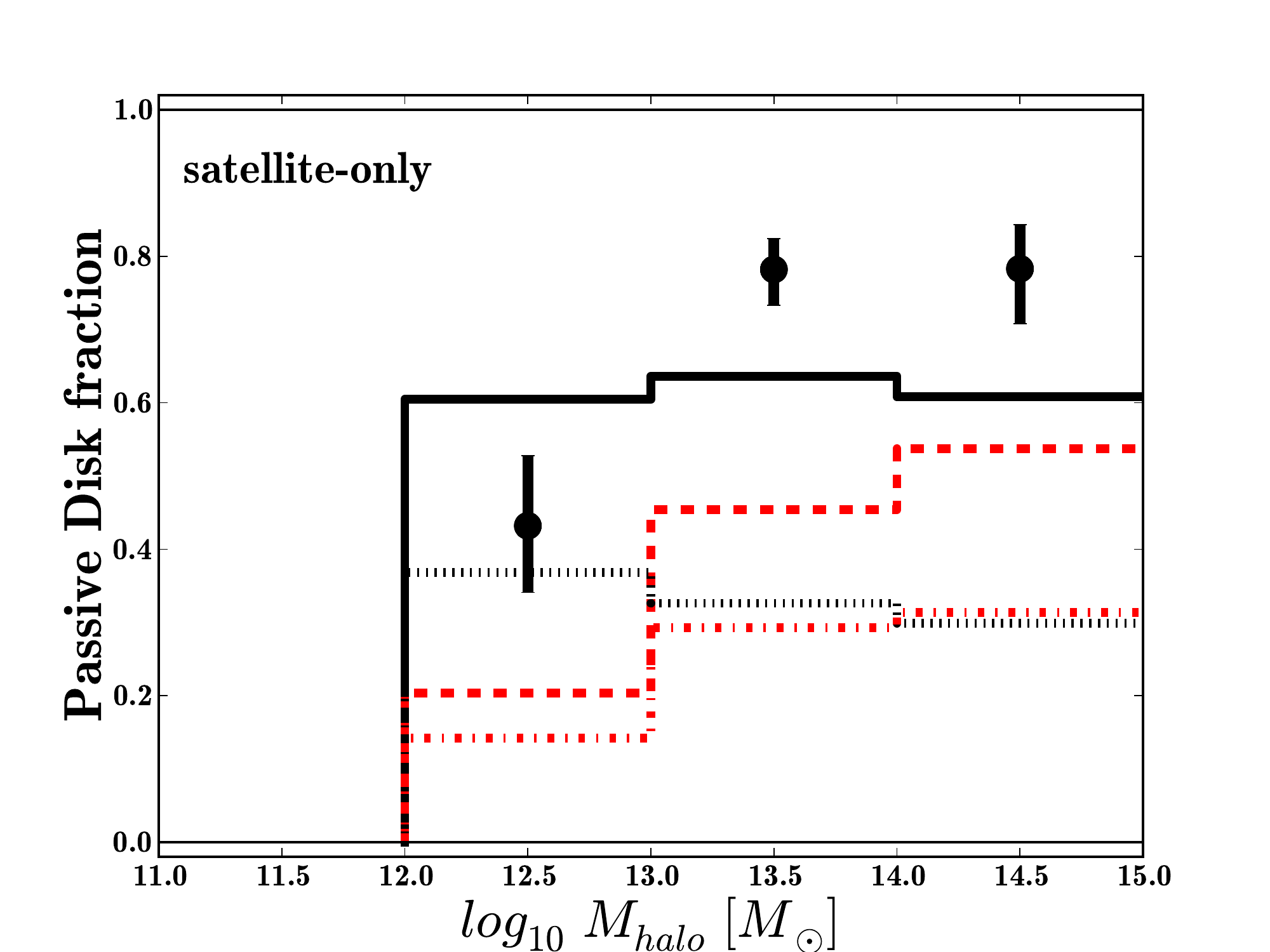}
            }
            \caption{Passive disc galaxy fraction. Otherwise as
              Figure~\ref{fig:fEz0std}.}
  \label{fig:fDpz0std}
\end{figure}

In the standard version of these models bulges are also formed when
discs become unstable and stars are transferred to the centre of a
galaxy. For consistency with the literature, we present here the
equivalent results for the \citetalias{WDL08} and \morgana\ models with
the {\it standard} bulge formation implementation in
figures~\ref{fig:fEz0std} (elliptical fraction),~\ref{fig:fDsfz0std}
(star-forming disc fraction) and~\ref{fig:fDpz0std} (passive disc
fraction). The pure mergers implementation is also shown for
comparison.

Figure~\ref{fig:fEz0std} clearly shows that disc instabilities add to
the fraction of both central and satellite ellipticals formed by the
\morgana\ model. While the \citetalias{WDL08} standard model does not
form many more central ellipticals, the fraction of satellite
ellipticals is enhanced by the inclusion of disc instabilities.  Disc
instabilities merely serve to increase the disparity between
observations and models, as the elliptical fraction was already too
high.

It is curious that \citetalias{WDL08} disc instabilities somehow lead
to a higher fraction of $\Mstellar\lesssim 10^{11.25}\Msol$ satellite
ellipticals, without contributing significantly to the formation of
central ellipticals.  Figure~\ref{fig:fDpz0std} shows that these are
otherwise passive discs in the pure merger model, and suggests that
passive discs are somehow more likely to suffer disc instabilities in
the \citetalias{WDL08} model. \morgana\ ellipticals formed via disc
instabilities do so in both central and satellite galaxies, and at the
expense of both passive and star-forming disc galaxies.

\end{document}